\documentclass[fleqn,usenatbib,useAMS]{mnras}

\usepackage[T1]{fontenc}
\usepackage{ae,aecompl}
\usepackage{float}
\usepackage{xcolor}

\usepackage{graphics,graphicx}	% Including figure files
\usepackage{amsmath}	% Advanced maths commands
\usepackage{amssymb}	% Extra maths symbols

\usepackage{newtxtext,newtxmath}

\usepackage{color,hyperref}
\hypersetup{colorlinks,breaklinks,
            linkcolor=blue,urlcolor=blue,
            anchorcolor=blue,citecolor=blue}
\title[The brightening of two Gaia-alerted young stars]{Photometric and spectroscopic study of the burst-like brightening of two \textit{Gaia}-alerted young stellar objects}

\author[Z. Nagy et al.]{
Zs\'ofia Nagy,$^{1,2}$\thanks{E-mail: nagy.zsofia@csfk.org}
P\'{e}ter \'{A}brah\'{a}m,$^{1,2,3}$
\'{A}gnes K\'{o}sp\'{a}l,$^{1,2,3.4}$ 
Sunkyung Park,$^{1,2}$
Micha{\l} Siwak,$^{1,2}$
\newauthor 
Fernando Cruz-S\'aenz de Miera,$^{1,2}$
Eleonora Fiorellino,$^{1,2,5}$
David Garc\'{i}a-\'{A}lvarez,$^{6,7}$
Zsófia Marianna Szab\'o,$^{1,2,8,9}$
\newauthor 
Simone Antoniucci,$^{5}$
Teresa Giannini,$^{5}$
Alessio Giunta,$^{10}$
Levente Kriskovics,$^{1,2}$
M\'aria Kun,$^{1,2}$
G\'abor Marton,$^{1,2}$
\newauthor 
Attila Mo\'or,$^{1,2}$
Brunella Nisini,$^{5}$
Andras P\'al,$^{1,2,3}$
L\'aszl\'o Szabados,$^{1,2}$
Pawe{\l} Zieli{\'n}ski,$^{11}$
{\L}ukasz Wyrzykowski$^{12}$
\\
$^{1}$Konkoly Observatory, Research Centre for Astronomy and Earth Sciences, E\"otv\"os Lor\'and Research Network (ELKH), \\ H-1121 Budapest, Konkoly Thege Mikl\'os \'ut 15-17., Hungary\\
$^{2}$CSFK, MTA Centre of Excellence, Budapest, Konkoly Thege Miklós út 15-17., H-1121, Hungary\\
$^{3}$ELTE E\"otv\"os Lor\'and University, Institute of Physics, P\'azm\'any P\'eter s\'et\'any 1/A, H-1117 Budapest, Hungary\\
$^{4}$Max Planck Institute for Astronomy, K\"onigstuhl 17, D-69117 Heidelberg, Germany \\
$^{5}$INAF-Osservatorio Astronomico di Roma, via di Frascati 33, 00078, Monte Porzio Catone, Italy \\
$^{6}$Instituto de Astrof\'{i}sica de Canarias, Avenida V\'{i}a L\'{a}ctea, E-38205 La Laguna, Tenerife, Spain 14 \\
$^{7}$Grantecan S.A., Centro de Astrof\'{i}sica de La Palma, Cuesta de San Jos\'{e}, E-38712 Bre\~{n}a Baja, La Palma, Spain \\
$^{8}$Max-Planck-Institut für Radioastronomie, Auf dem Hügel 69, 53121 Bonn, Germany \\
$^{9}$Scottish Universities Physics Alliance (SUPA), School of Physics and Astronomy, University of St Andrews, North Haugh, St Andrews, KY16 9SS, UK \\
$^{10}$Space Science Data Center, Italian Space Agency, via del Politecnico, 00133, Roma, Italy \\
$^{11}$Institute of Astronomy, Faculty of Physics, Astronomy and Informatics, Nicolaus Copernicus University in Toru{\'n},\\ ul. Grudzi\k{a}dzka 5, 87-100 Toru{\'n}, Poland \\
$^{12}$Astronomical Observatory, University of Warsaw, Al. Ujazdowskie 4, 00-478, Warsaw, Poland \\
}
\date{Accepted XXX. Received YYY; in original form ZZZ}

\pubyear{2022}
 
\begin{document}
\label{firstpage}
\pagerange{\pageref{firstpage}--\pageref{lastpage}}
\maketitle

\begin{abstract}

Young stars show variability on diﬀerent time-scales from hours to decades, with a range of amplitudes. We studied two young stars, which triggered the Gaia Science Alerts system due to brightenings on a time-scale of a year. Gaia20bwa brightened by about half a magnitude, whereas Gaia20fgx brightened by about two and half magnitudes. We analyzed the Gaia light curves, additional photometry, and spectra taken with the Telescopio Nazionale Galileo and the Gran Telescopio Canarias. Several emission lines were detected toward Gaia20bwa, including hydrogen lines from H$\alpha$ to H$\delta$, Pa$\beta$, Br$\gamma$, and lines of \ion{Ca}{ii}, \ion{O}{i}, and \ion{Na}{i}. The H$\alpha$ and Br$\gamma$ lines were detected toward Gaia20fgx in emission in its bright state, with additional CO lines in absorption, and the Pa$\beta$ line with an inverse P Cygni proﬁle during its fading. Based on the Br$\gamma$ lines the accretion rate was $(2.4-3.1)\times10^{-8}$ $M_\odot$ yr$^{-1}$ for Gaia20bwa and $(4.5-6.6)\times10^{-8}$ $M_\odot$ yr$^{-1}$ for Gaia20fgx during their bright state. The accretion rate of Gaia20fgx dropped by almost a factor of 10 on a time-scale of half a year. The accretion parameters of both stars were found to be similar to those of classical T Tauri stars, lower than those of young eruptive stars. However, the amplitude and time-scale of these brightenings place these stars to a region of the parameter space, which is rarely populated by young stars. This suggests a new class of young stars, which produce outbursts on a time-scale similar to young eruptive stars, but with smaller amplitudes.

\end{abstract}


\section{Introduction} 

About half of young stellar objects (YSOs) show photometric variations on daily-weekly timescales, with typical values of a few times 0.1 mag at optical and infrared wavelengths (e.g. \citealp{Megeath2012}). 
Some young stars show brightness variations on even longer time-scales: months, years, centuries. These variations are related to different physical processes \citep{HillenbrandFindeisen2015}. Some of the light variations are periodic, and are related to photospheric inhomogeneities, such as starspots. 
Periodic or quasi-periodic dips in the light curves can be related to circumstellar dust passing through the line of sight toward the star. One example of this phenomenon is the class of AA Tau-type stars, where the occultations are caused by an inner disk warp (e.g. \citealp{Bouvier1999,Bouvier2003}, \citealp{Cox2013}, \citealp{Nagy2021}).
Aperiodic events can also occur due to the unsteady mass transfer from the inner disk to the star \citep{blinova2016}, including ''clumpy accretion'' \citep{gullbring1996,siwak2018}.
The eruptive class of young stars shows brightness variations with an amplitude of a few magnitudes and remain bright on longer timescales. These events are typically called outbursts and are caused by a sudden increase of the mass accretion rate from $10^{-10}-10^{-8}$ in quiescence up to $10^{-6}-10^{-4}$ M$_\odot$ yr$^{-1}$ during outburst. 
Eruptive young stars are commonly divided into two main classes: EX Lupi-type stars (EXors) and FU Orionis-type stars (FUors). 
The former show brightenings of 2-4 mag, last for less than a year and are recurrent (e.g. \citealp{Herbig2008}); the latter brighten by up to 5 magnitudes and last for several decades (e.g. \citealp{Audard2014}). So far the number of confirmed FUors is limited to no more than a dozen while the number of known EXors is limited to less than 25, including candidates \citep{Audard2014}. 
Recent studies have shown, that not all the eruptive young stars belong to these two main classes (\citealp{Hillenbrand2019}, \citealp{Hodapp2020}).

ESA's \textit{Gaia} space telescope has been monitoring the whole sky, determining the parallax and proper motion of 1.8 billion stars. The sources which show significant brightness changes are reported as \textit{Gaia} Photometric Science Alerts \citep{Hodgkin2021}, which are well suited to identify brightening or fading events of young stars. 
Two \textit{Gaia} Science alert sources have already been identified and confirmed as FUors: Gaia17bpi \citep{Hillenbrand2018} and Gaia18dvy \citep{SzegediElek2020}, two other \textit{Gaia} alert sources as EXors: Gaia18dvz \citep{Hodapp2019} and Gaia20eae \citep{CruzSaenzdeMiera2022}, while two more sources were found to be young eruptive stars other than FUors or EXors:  Gaia19ajj \citep{Hillenbrand2019} and Gaia19bey \citep{Hodapp2020}. Another young star, V555~Ori was also reported as a \textit{Gaia} alert due to a brightening event, but was later found to be a result of changing circumstellar extinction rather than an accretion burst \citep{Nagy2021}.
\textit{Gaia} Science alerts were also published for the two targets analyzed below as well.

Gaia20bwa, or 05351885-0444100 in the Two Micron All Sky Survey (2MASS) All-Sky Catalog of Point Sources \citep{cutri2003} ($\alpha_{\rm J2000}$ = 05$^{\rm h}$ 35$^{\rm m}$ 18$\fs$86, $\delta_{\rm J2000}$ = $-$4$^{\circ}$ 44$'$ 10$\farcs$21), is located in the Orion A star-forming region at 410$^{+12}_{-11}$ pc \citep{BailerJones2021}. 
\citet{Tobin2009} found it to be a classical T Tauri star (CTTS), but didn't rule out the possibility of being a weak-line T Tauri star. 
\citet{DaRio2016} determined the effective temperature of Gaia20bwa to be 3142.5$\pm$16.0~K and its mass to be 0.206$\pm$0.008 M$_\odot$.
Its brightening by 0.3 mag was reported as a \textit{Gaia} alert on 2020 April 18. Gaia20bwa is part of the catalogue of \citet{Marton2019} with an 85\% probability of being a YSO.
Gaia20fgx ($\alpha_{\rm J2000}$ = 22$^{\rm h}$ 54$^{\rm m}$ 59$\fs$09, $\delta_{\rm J2000}$ = $+$62$^{\circ}$ 34$'$ 34$\farcs$75) is located in the Cepheus OB3 association, at about 1.01$^{+1.88}_{-0.32}$ kpc \citep{BailerJones2021}. 
It had a \textit{Gaia} alert on 2020 November 17 due to a 2.5 mag brightening over a year.
Gaia20fgx is also part of the catalogue of \citet{Marton2019} with a 96\% probability of being a YSO. \citet{Allen2012} identified this target as a Class II source.

In this paper, we analyze follow-up photometry and spectroscopy of the sources in order to investigate whether their brightening is related to eruptive events or scaled-up version of the normal magnetospheric accretion.
In Section \ref{sect_observations} we describe the photometric and spectroscopic observations of the sources, in Sect. \ref{sect_results} we explain the results on the light curves, color variations, spectroscopy, spectral energy distributions (SEDs), and accretion rates of the targets. We compare the results to those toward eruptive young stars and CTTS in Sect. \ref{sect_discussion}.

\section{Observations}   
\label{sect_observations}

\subsection{Photometry}

\begin{figure}
\centering
\includegraphics[width=0.47\columnwidth]{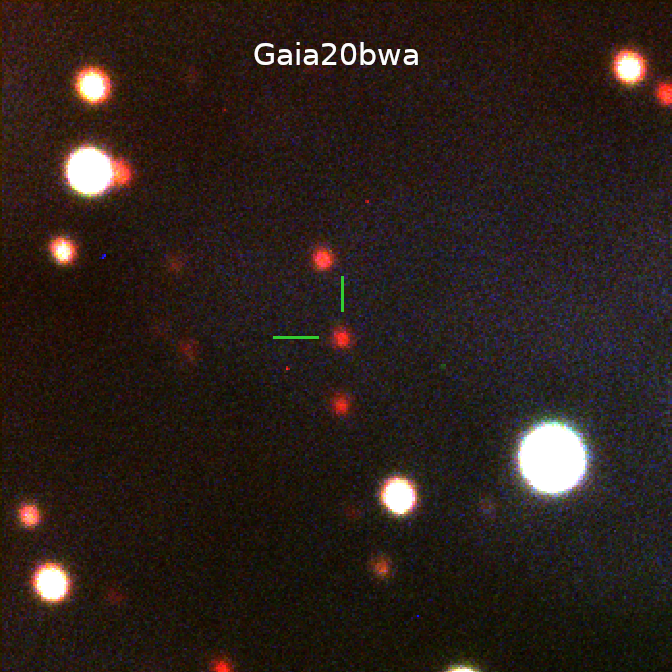}
\includegraphics[width=0.47\columnwidth]{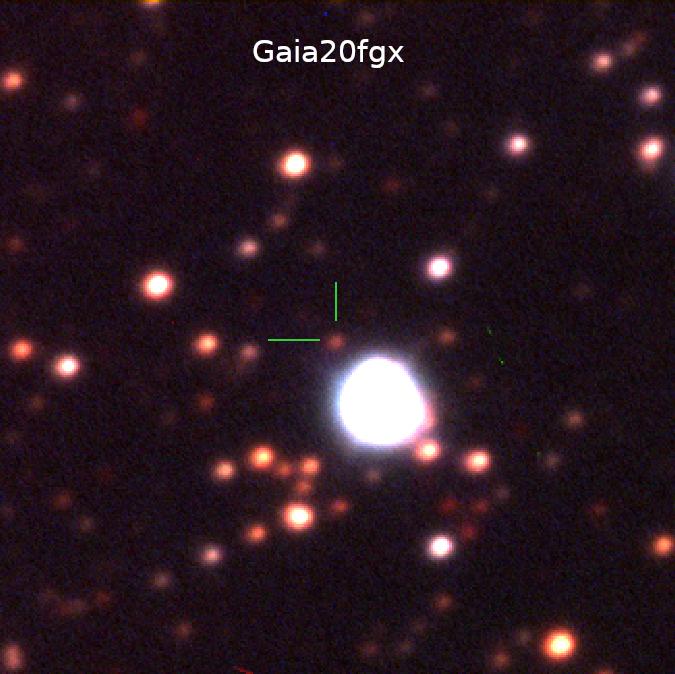}
\caption{$Vr'i'$ color composite images of Gaia20bwa (left) and Gaia20fgx (right). The images show $3'{\times}3'$ areas centered on the targets. North is up, east is to the left.}
\label{fig:findingcharts}
\end{figure}

We obtained ground-based optical photometric observations of Gaia20bwa and Gaia20fgx with the 80\,cm Ritchey-Chr\'etien telescope (RC80) at the Piszk\'estet\H{o} Mountain Station of Konkoly Observatory (Hungary) and with the 60\,cm Carl-Zeiss telescope at Mount Suhora Observatory of the Cracow Pedagogical University (Poland). The RC80 telescope was equipped with an FLI PL230 CCD camera, 0$\farcs$55 pixel scale, $18\farcm8\times18\farcm8$ field of view, Johnson $BV$ and Sloan $g'r'i'$ filters. The telescope at Mount Suhora was equipped with an Apogee Aspen-47 camera, 1$\farcs$116 pixel scale, $19\farcm0\times19\farcm0$ field of view, Sloan $g'r'i'$ filters. \autoref{fig:findingcharts} shows portions of the images taken of our targets. We typically obtained 3 to 13 images in each filter. We first applied CCD reduction including bias, flatfield, and dark current corrections. Then we calculated aperture photometry for the science target and several comparison stars in the field of view using an aperture radius of 2$\farcs$75. 
We selected those comparison stars from the APASS9 catalog \citep{henden2015} that were within 6$\farcm$5 of the target and which were mostly constant, i.e., the rms of their V-band observations from the ASAS-SN Photometry Database \citep{shappee2014, jayasinghe2019} were below 0.1 mag. The APASS9 catalog provided Bessell $BV$ and Sloan $g'r'i'$ magnitudes for the comparison stars. We used the comparison stars for the photometric calibration by fitting a linear color term. Magnitudes taken with the same filter on the same night were averaged. The final uncertainties are the quadratic sum of the formal uncertainties of the aperture photometry, the photometric calibration, and the scatter of the individual magnitudes that were averaged per night. 
The results can be found in Tab.~\ref{tab:phot} in the Appendix.
As seen in Fig. \ref{fig:findingcharts}, Gaia20fgx is close to a very bright star, however, given the used aperture, and that the separation between Gaia20fgx and the bright star is $\sim$7$''$, the results from the photometry were not affected by this nearby source.

 We obtained near-infrared photometric observations in the $J$, $H$ and $K_{\rm s}$ bands of Gaia20bwa on 2021 February 11 and Gaia20fgx on 2021 January 27. We used the Near Infrared Camera Spectrometer (NICS) instrument on the Telescopio Nazionale Galileo (TNG) located in the Island of San Miguel de La Palma (Canary Islands, Spain), proposal ID: AOT42, PI: Eleonora Fiorellino. In each filter a 5-point dithering was performed, with 3~s of exposure time at each position. The data reduction, performed with our own {\sc IDL} routines, included the construction and subtraction of a sky image, and flat-fielding. 
On 2021 September 18, we obtained additional $JHK_{\rm s}$ photometry of Gaia20fgx from the Gran Telescopio Canaria (GTC) using the Espectr\'ografo Multiobjeto Infra-Rojo (EMIR) instrument \citep{Balcells2000}, proposal ID: GTC01-21BDDT, PI: David Garc\'{i}a. The images were obtained in four dither positions with 10$''$ offsets with 5~s exposure per dither point. These data were processed using the PyEMIR \citep{Pascual2010} pipeline version 0.163.
For photometric calibration of the TNG and GTC photometry, the 2MASS catalog was adopted. We extracted the instrumental magnitudes for the target as well as for all good-quality 2MASS stars (i.e. with a 2MASS photometric quality flag of AAA) in the field in an aperture with a radius of $\sim$1.5$''$. 
The final step was the determination of an average constant calibration factor between the instrumental and the 2MASS magnitudes of typically $30-50$ stars, and this offset was applied on the target observations. The formal photometric uncertainties are $0.01-0.02$ mag for the TNG data and $0.07-0.14$ for the GTC data. The results can be found in Tab.~\ref{tab:phot} in the Appendix.

We also used mid-infrared photometry from the Wide-field Infrared Survey Explorer (\textit{WISE}) and \textit{NEOWISE} surveys from the NASA/IPAC Infrared Science Archive. \textit{NEOWISE} observes the full sky on average twice per year with multiple exposure per epoch. For a comparison with the photometry from other instruments, we computed the average and standard deviation of multiple exposures of a single epoch. The error bars are a quadratic sum of the average magnitude uncertainty per exposure.
We downloaded $G$ band photometry from the \textit{Gaia} Science Alerts Index website. 
We also used $r$ and $g$ band photometry from the Data Release 11 of the Zwicky Transient Facility (ZTF; \citealp{Masci2019}).

\subsection{Optical spectroscopy}

We used the TNG equipped with the Device Optimized for the LOw RESolution (Dolores) to obtain low-resolution optical spectra of Gaia20bwa on 2021 February 11/12, and Gaia20fgx on 2021 January 27/28. The UV-NIR coverage was achieved using the LRB and LRR gratings, which operate in $\sim3500-7980$ and $\sim4980-10370$~\AA~spectral ranges, respectively.
The spectra were reduced in a standard way on bias, flatfield and then wavelength-calibrated within {\sc IRAF}.
The strong fringing pattern (apparent above 7800~\AA) was successfully removed by normalized flatfield division. This worked only for the Gaia20bwa spectrum, the same procedure failed for the second star. This appears to be have been caused by the low elevation at which the telescope was pointing during the observation of Gaia20fgx, at an airmass range $2.5-2.7$, resulting in increased instrumental flexures distorting the nominal optical path. 
We averaged the overlapping part of LRB and LRR ranges assuming weights appropriate to the obtained signal. 
In addition, we observed Gaia20bwa on 2021 February 17/18 using the 2-m Liverpool Telescope (LT) equipped with the SPectrograph for the Rapid Acquisition of Transients (SPRAT, \citealp{Piascik2014})~\footnote{ProgID: XOL20B01, PI: Pawel Zielinski}. SPRAT provides low-resolution (R=350) spectra in the range of $4020-7994$~\AA. The spectrum was reduced and approximately calibrated to absolute flux units by means of the dedicated SPRAT pipeline.
The spectral resolutions and total integration times corresponding to the optical spectra are listed in Table \ref{tab:spec}.

\subsection{NIR spectroscopy}

Low resolution (R=500-1250) NIR spectra were obtained with the Near Infrared Camera Spectrometer (NICS, \citealp{Baffa2001}) installed on the TNG on 2021 February 11 for Gaia20bwa and on 2021 January 27 for Gaia20fgx.
For Gaia20bwa the $J$, $HK$, and $K_b$ bands were used with exposure times of 1000 s, 120 s, and 520 s, respectively. For Gaia20fgx, the $IJ$, $HK$, and $K_b$ bands were used with exposure times of 2000 s, 400 s, and 1400 s, respectively. The sources were observed through the $1''$ wide slit.
The data were reduced using {\sc IRAF}. For each image, sky subtraction, flat-fielding, bad pixel removal, aperture tracing, and wavelength calibration (using argon lamp) were performed. 
Then, the telluric correction was performed: the hydrogen absorption lines in the telluric stellar spectrum were removed by Gaussian fitting, and the telluric spectrum was normalized. 
Subsequently we divided the target spectrum by the normalized telluric spectrum. The barycentric velocity was calculated by barycorrpy \citep{kanodia2018} as --23.22 and --15.14 km~s$^{-1}$ for Gaia20bwa and Gaia20fgx, respectively, and then subtracted from the target spectra. Finally, after normalizing the telluric corrected target spectrum, flux calibration was performed by using our photometry from Mt. Suhora and the TNG during the bright state, and the \textit{WISE} W1 and W2 photometry close to the bright state for Gaia20bwa.
For Gaia20fgx, faint states of \textit{WISE} data are used in addition to bright states of optical to NIR.

We obtained medium-resolution (R=4000–5000) spectra for Gaia20fgx in $JHK_{\rm s}$-bands on 2021 September 19, using the GTC equipped with EMIR configured in the long-slit mode (PI: D. Garc\'{i}a). 
The star was observed through the $0\farcs8$ wide slit. The total exposure times were 7174~s in the $J$, 2795~s in the $H$, and 3235~s in the $Ks$ band. HgAr lamp provided wavelength calibration. The spectra were obtained in ABBA nodding pattern along the slit and were processed by means of the dedicated PyEMIR package. The final spectrum extraction was performed under IRAF. The telluric correction was done using the telluric standard HD 212495.
We calibrated these spectra to absolute flux using the $JHK_{\rm s}$ band photometry obtained on the same night.
The spectral resolutions and total integration times corresponding to the NIR spectra are listed in Table \ref{tab:spec}.

\begin{table*}
\centering
\caption{Log of spectroscopic observations.}
\label{tab:spec}
\begin{tabular}{ccccccc}
\hline \hline
Target  &  Date & Telescope& Instrument& Wavelength ($\mu$m)& Resolution & Total int. time (s)\\
\hline
Gaia20fgx& 2021 Jan 27& TNG& DOLORES& 0.59--1.00& 585--714& 1700\\
%800(LRB) \& 900(LRR)
Gaia20fgx& 2021 Jan 27& TNG& NICS&    0.90--1.45& 500& 8000\\ 
Gaia20fgx& 2021 Jan 27& TNG& NICS&    1.40--2.50& 500& 1600\\ 
Gaia20fgx& 2021 Jan 27& TNG& NICS&    1.95--2.34& 1250& 5600\\ 
Gaia20fgx& 2021 Sep 19& GTC& EMIR&    1.17--1.33& 4000--5000& 7174\\
Gaia20fgx& 2021 Sep 19& GTC& EMIR&    1.53--1.78& 4000--5000& 2795\\
Gaia20fgx& 2021 Sep 19& GTC& EMIR&    2.03--2.38& 4000--5000& 3235\\
\hline
Gaia20bwa& 2021 Feb 17& LT&  SPRAT&   0.40--0.80& 350& 200\\
Gaia20bwa& 2021 Feb 11& TNG& DOLORES& 0.59--1.00& 585--714& 630\\
%280(LRB) \& 350(LRR)
Gaia20bwa& 2021 Feb 11& TNG& NICS&    1.12--1.40&          1200&    4000\\ 
Gaia20bwa& 2021 Feb 11& TNG& NICS&    1.40--2.50&          500&     480\\ 
Gaia20bwa& 2021 Feb 11& TNG& NICS&    1.95--2.34&          1250&    2080\\ 
\hline
\end{tabular}
\end{table*}

\section{Results}
\label{sect_results}

\subsection{The distance of Gaia20fgx}
\label{sect_distance}

\begin{figure}
\centering
\includegraphics[width=9.5cm, trim={1.5cm 0 0 1cm},clip]{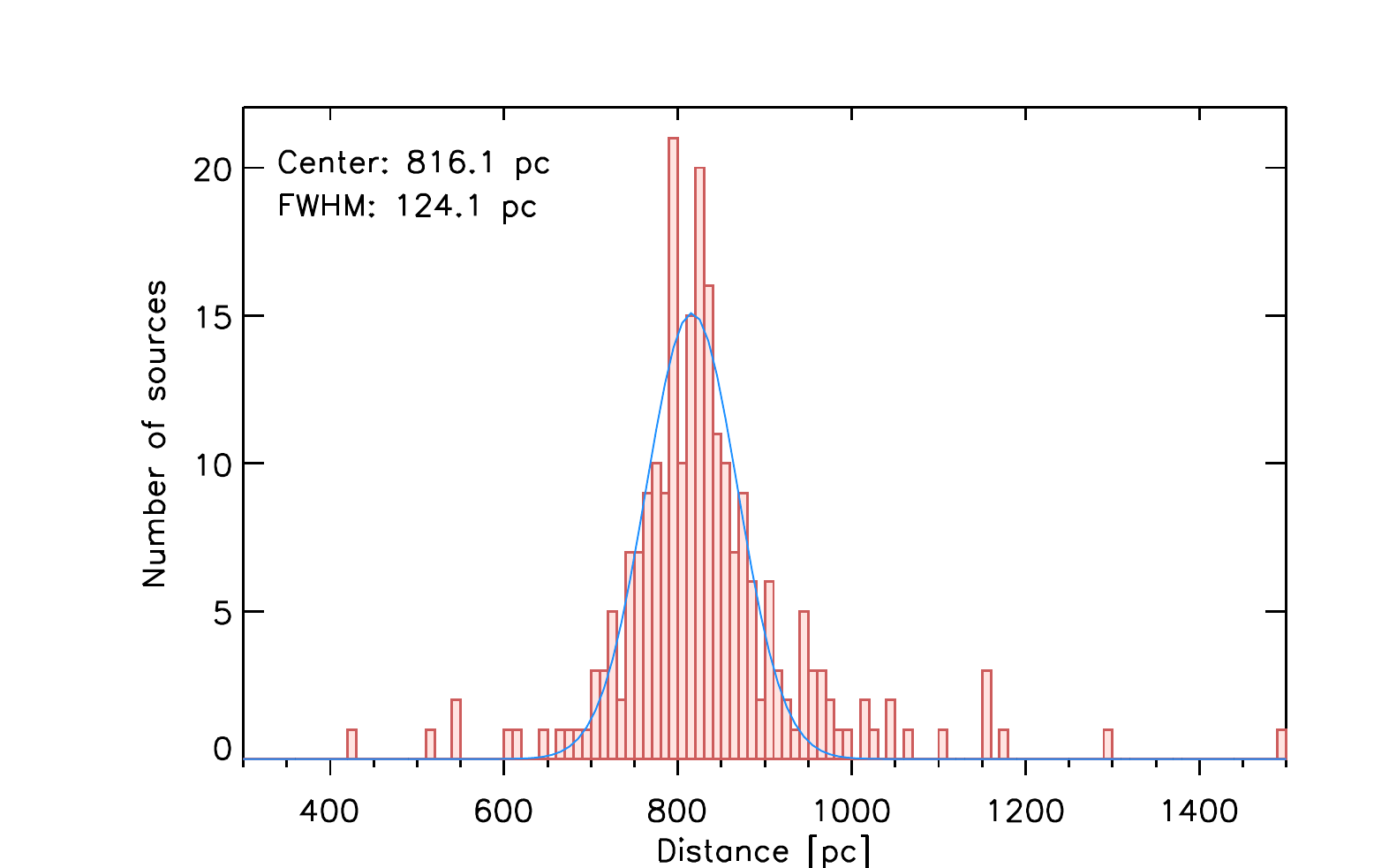}
\caption{Histogram of \textit{Gaia} EDR3 distances of the members of the Cep OB3 region in 10 pc bins. The fitted Gaussian peaks at a value of $\sim$816 pc.}
\label{fig:distance}
\end{figure}

While the distance of Gaia20bwa is accurately known, the distance of Gaia20fgx (1.01$^{+1.88}_{-0.32}$ kpc, \citealp{BailerJones2021}) has a large uncertainty.
Based on its position, proper motion, and its distance, Gaia20fgx belongs to the Cep OB3 association. To derive its more accurate distance, we collected the list of sources, which also belong to the Cep OB3 association (\citealp{Jordi1996}, \citealp{Getman2009}), and downloaded their Gaia EDR3 distances from \citet{BailerJones2021}. We considered only those sources, that had a Renormalised Unit Weight Error (RUWE) below 1.4. We removed the sources with a negative parallax from our sample, and those, where the parallax was less than five times its error. After applying these selection criteria, the number of sources used for the analysis is 235.
The number or sources per photogeometric distance in 10 pc bins is shown in Fig. \ref{fig:distance}. 
Based on a Gaussian fit, the distribution of distances peaks at $\sim$816 pc. The FWHM of the Gaussian fit is $\sim$124 pc. We use the Gaussian sigma of $\sigma=FWHM / 2.355 = 52.7$ pc as the error of the distance.
Based on this analysis, the distance of Cep OB3 is $\sim$816$\pm$53 pc. We will use this value when calculating physical parameters for Gaia20fgx in the next sections.

\subsection{Light curves}

\begin{figure*}
\centering
\includegraphics[width=\textwidth]{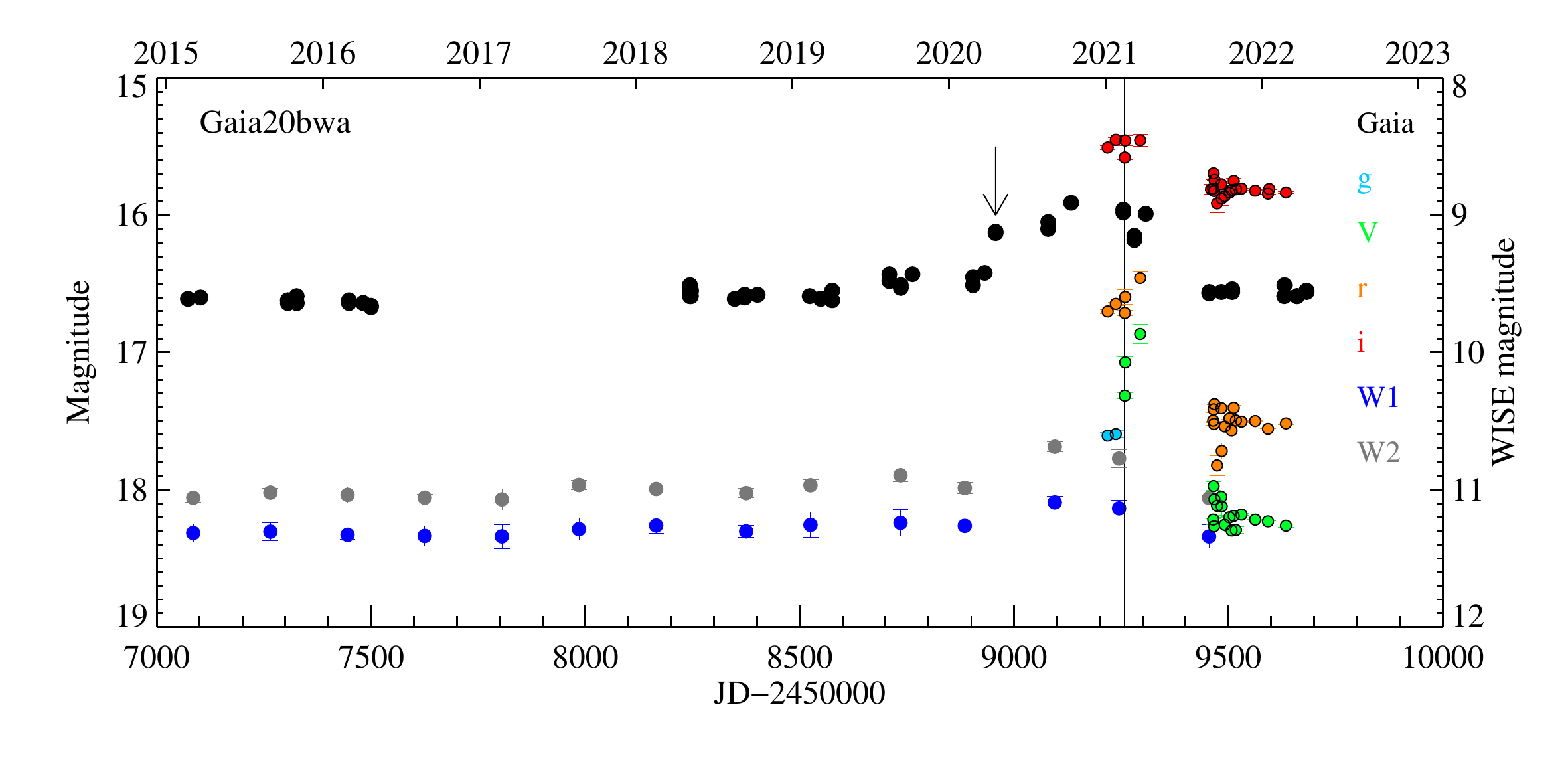}
\includegraphics[width=\textwidth]{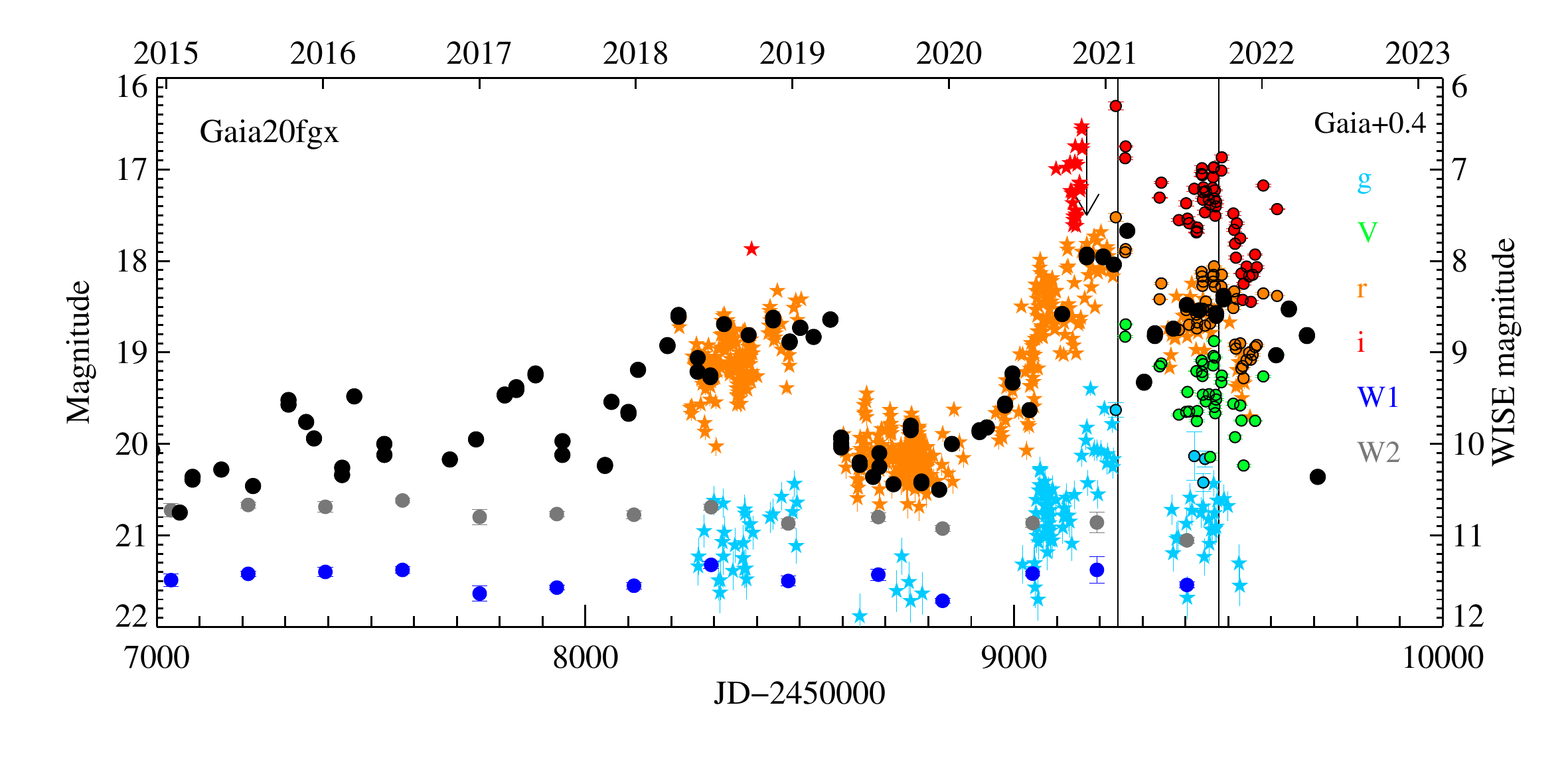}
\caption{Light curves of Gaia20bwa and Gaia20fgx. The \textit{Gaia} G data are shown with black dots, the ZTF $g$, $r$, and $i$ data with light blue, orange, and red asterisks, the \textit{WISE} W1 and W2 band data with blue and grey dots.
The red, orange, and green dots are $i$, $r$, and $V$ data measured with the RC80 or the Mt. Suhora telescope, as shown in Appendix A.
The vertical lines show the epochs of the TNG and GTC spectra.
The arrows show the epochs of the \textit{Gaia} alerts.
}
\label{fig:light}
\end{figure*}

\textit{Gaia} and \textit{WISE} light curves of Gaia20bwa and Gaia20fgx are shown in Fig. \ref{fig:light}. For Gaia20fgx we also show data from the Zwicky Transient Facility (ZTF) archive\footnote{\hyperlink{https://irsa.ipac.caltech.edu/Missions/ztf.html}{https://irsa.ipac.caltech.edu/Missions/ztf.html}}. 
For Gaia20bwa, we did not consider the ZTF data, due to the quality of the photometry in the ZTF Data Release 11: most of the data points are defined as bad-quality.

Gaia20bwa had a \textit{Gaia} alert on 2020 April 17, when its brightness increased by about 0.3 mag in the \textit{Gaia} $G$-band. It continued brightening by about 0.2 mag to reach its maximum brightness in 2020 October.
The last \textit{WISE} data point also follows the brightening seen in the \textit{Gaia} light curve.
Based on the \textit{Gaia} light curve as well as our follow-up photometry with the RC80, Gaia20bwa faded back to its long-term brightness by the end of 2021 August, therefore, the brightening episode lasted for approximately 17 months.

Gaia20fgx had a \textit{Gaia} alert on 2020 November 11, due to its brightening by $\sim$2.5 mag over about 10 months. It reached its maximum brightness by 2021 February, and returned to quiescence by 2022 May, based on the $G$-band data. The \textit{Gaia} light curve also shows an earlier brightening from early 2018 until early 2019, with a lower amplitude compared to the second brightening that corresponds to the \textit{Gaia} alert. In addition to the two long-term brightening events, shorter brightenings with an amplitude of $\sim$1 mag are apparent on the \textit{Gaia} light curve between 2015 and 2018.

\subsection{Color variations}

\begin{figure}
\centering
\includegraphics[width=9cm, trim={0cm 0 0 2cm},clip]{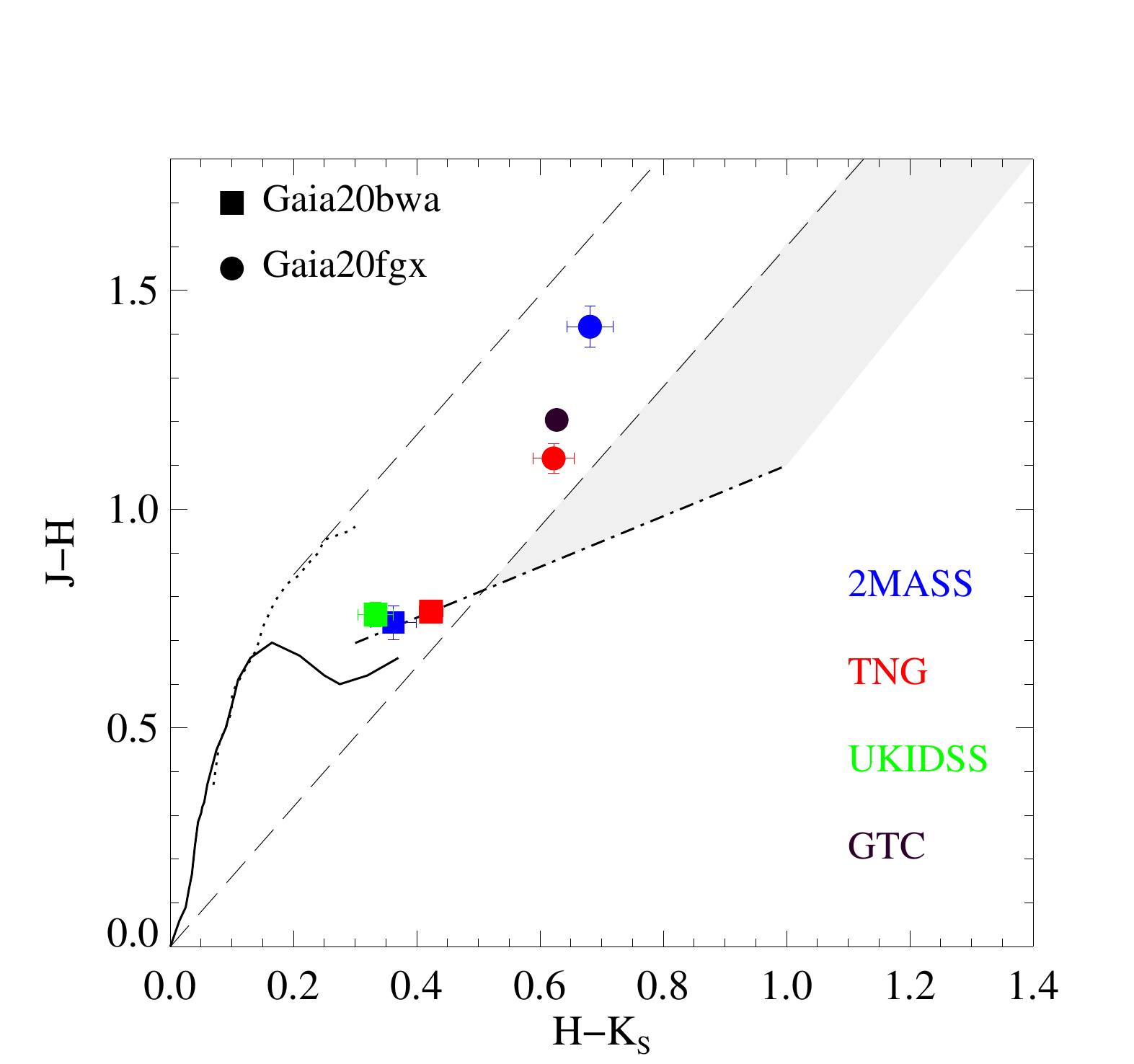}
\caption{($J-H$) versus ($H-K_S$) color–color diagram. 
The filled circles correspond to Gaia20fgx, the squares to Gaia20bwa. The blue symbols show 2MASS data, the red points TNG data, and the green symbol is from the UKIRT Infrared Deep Sky Survey (UKIDSS).
The solid curve shows the colors of the zero-age main-sequence, and the dotted line
represents the giant branch \citep{BessellBrett1988}. The long-dashed lines
delimit the area occupied by the reddened normal stars \citep{Cardelli1989}.
The dash–dotted line is the locus of unreddened CTTS \citep{Meyer1997} and the grey shaded band borders the area of the reddened $K_S$-excess stars.}
\label{fig:infra_color}
\end{figure}

Figure \ref{fig:infra_color} shows the ($J-H$) versus ($H-K_S$) color--color diagram of the sources. Gaia20bwa is close to the locus of unreddened CTTS based on each three data points, and in both the bright and faint states. This suggests that the visual extinction for Gaia20bwa is low, however, we will use the SEDs in Sect. \ref{sect:sed} to constrain the $A_V$ for Gaia20bwa.
For Gaia20fgx, there is evidence for a change in the visual extinction between the faint state represented by the 2MASS data point and the bright state represented by the TNG data point, while the GTC photometry corresponds to the fading of the source, between the bright and faint states. Gaia20fgx was redder at the 2MASS epoch, than at the later epochs observed with the TNG and the GTC.
For Gaia20fgx, we used the expression from \citet{Cardelli1989} to measure the visual extinction of the source at each epoch by projecting its location in Fig. \ref{fig:infra_color} to the line representing the locus of unreddened CTTS \citep{Meyer1997} along the extinction path. This method results in an $A_V = 3.6 \pm 0.2$ mag for the bright state and $5.7 \pm 0.6$ mag for the faint state, when assuming an $R_V$ of 3.1. An $A_V = 4.1 \pm 0.2$ mag was found for the GTC data point. 

\begin{figure*}
\centering
\includegraphics[width=4.4cm]{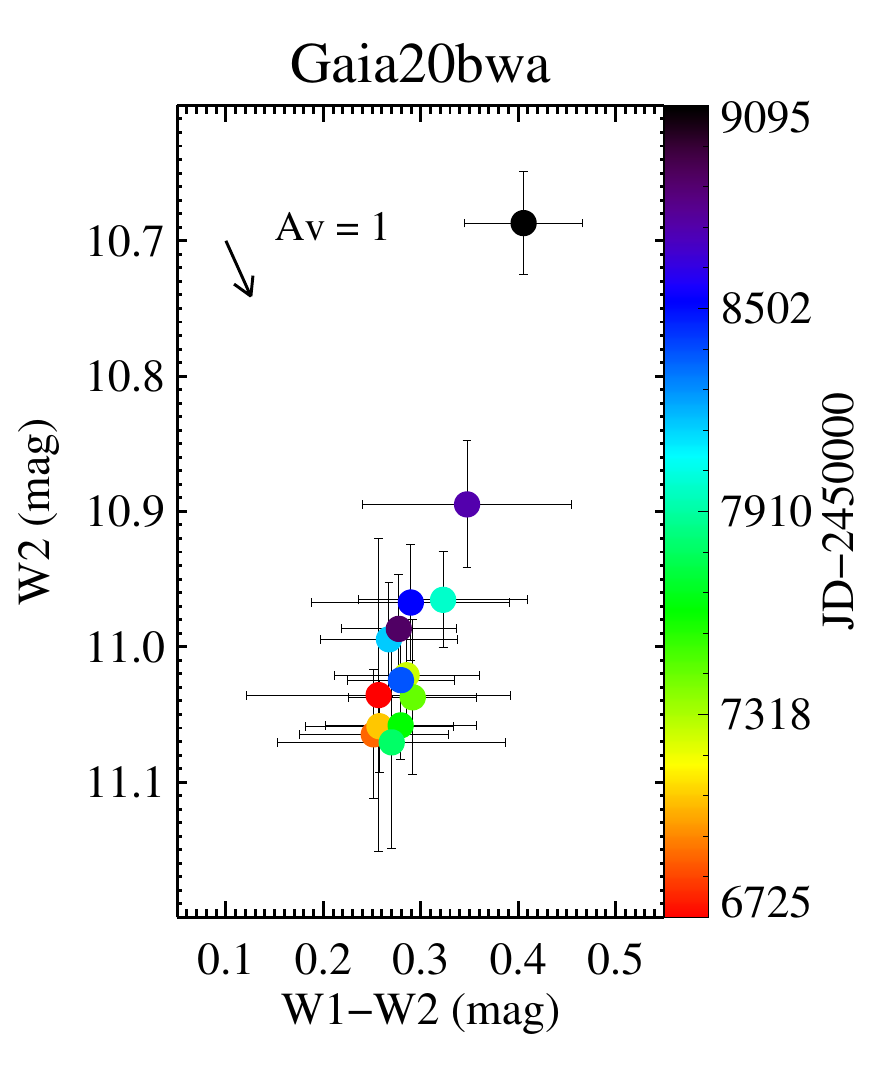}
\includegraphics[width=4.4cm]{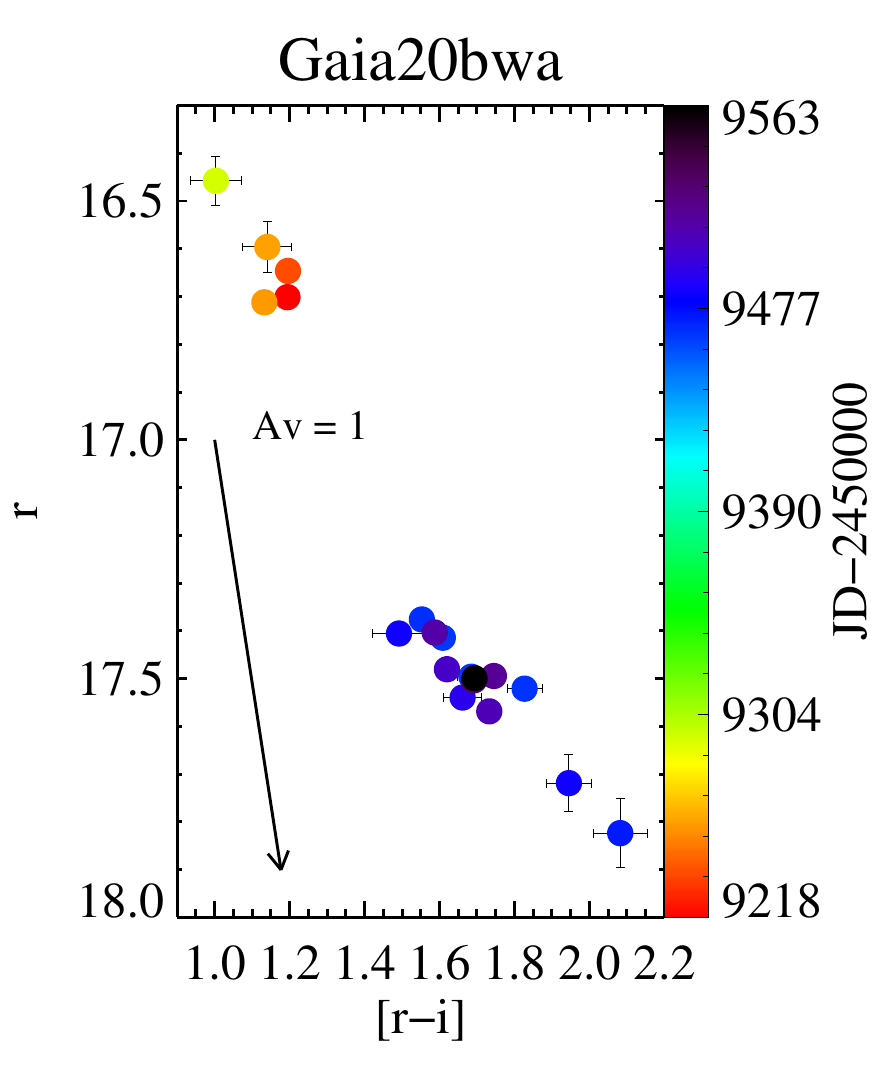}
\includegraphics[width=4.4cm]{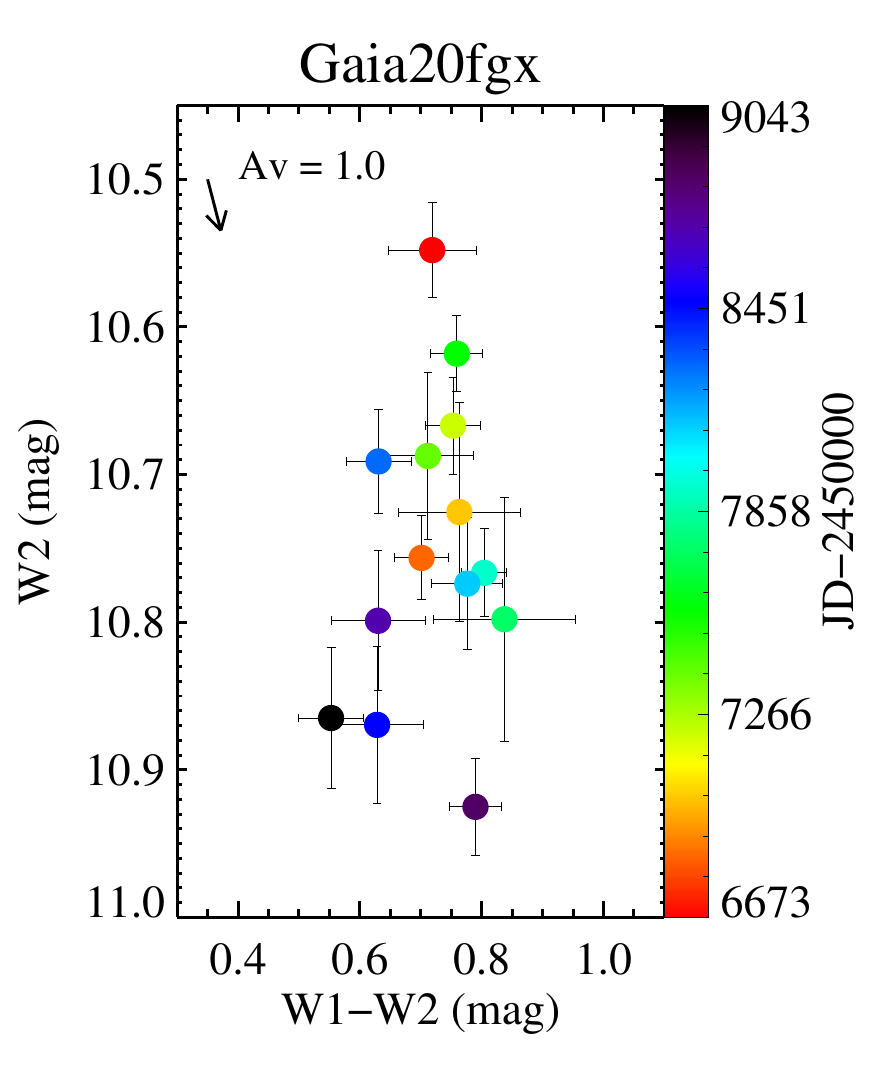}
\includegraphics[width=4.4cm]{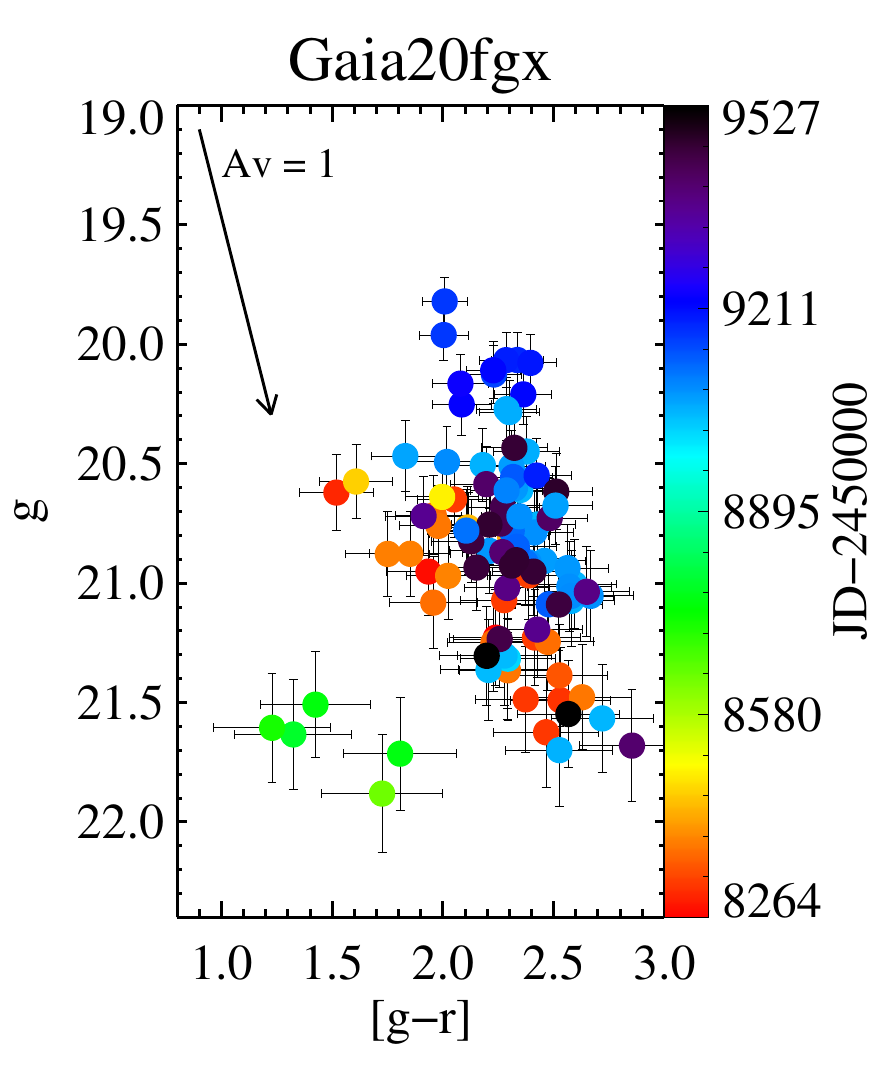}
\caption{\textit{Left panels}: Color--magnitude diagrams of Gaia20bwa based on data from \textit{WISE} and from the RC80 telescope. 
\textit{Right panels}: Color--magnitude diagrams of Gaia20fgx based on data from \textit{WISE} and follow-up photometry in $g$ and $r$ bands.
The $g$ versus [$g-r$] color--magnitude diagrams are based on data from the ZTF survey as well as from the RC80 and Mt. Suhora telescopes as shown in Table A.1. We only used those ZTF data points, when both $g$ and $r$ magnitudes were measured during the same night.}
\label{fig:colormag_wise}
\end{figure*}

Color--magnitude diagrams based on the \textit{WISE} data are shown in Fig. \ref{fig:colormag_wise}. 
There is no significant color change in the case of Gaia20fgx. For Gaia20bwa, the W2 versus W1--W2 color--magnitude diagram suggests reddening during the brightening, which is likely due to the disk component.

Figure \ref{fig:colormag_wise} also shows a $g$ vs. [$g-r$] color--magnitude diagram for Gaia20fgx based on archival data from the ZTF survey and our follow-up observations with the RC80 and Mt. Suhora telescopes. Some of the data points indicate color-variations related to changing extinction. 
Both brightening events seen in Fig. \ref{fig:colormag_wise} based on the $g$ vs. [$g-r$] color--magnitude diagram show a linear color variation over time: the red-orange data points correspond to the first brightening event, while the blue data points to the brightening related to the \textit{Gaia} alert. The data points corresponding to the second brightening event seem to follow the extinction path.
The $r$ versus [$r-i$] color--magnitude diagram shown in Fig. \ref{fig:colormag_wise} for Gaia20bwa based on data points during the bright state and the fading shows a linear trend, however, it is not consistent with the extinction path. It suggests that the fading was caused by a mechanism other than variable extinction.

\subsection{Results from spectroscopy}

Due to the low spectral resolution of the TNG and LT spectra, the velocities of the lines cannot be accurately determined. However, the equivalent widths and fluxes of the lines can be measured and used as tracers of characteristics of the sources, such as the accretion rate.

\begin{figure*}
\centering
\includegraphics[width=\textwidth]{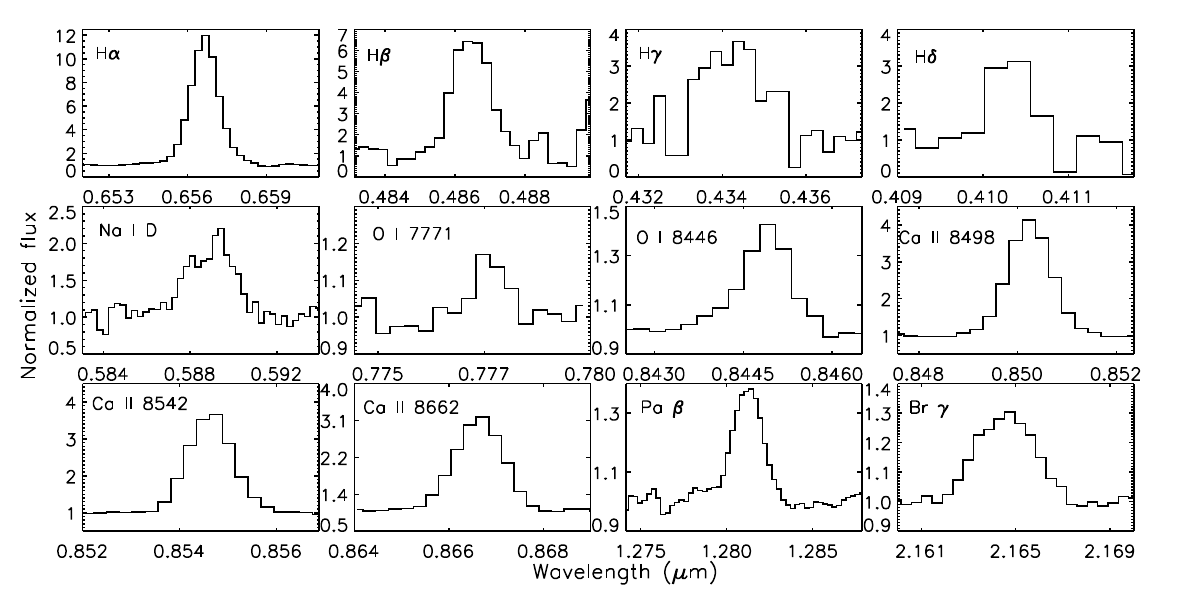}
\caption{Lines detected toward Gaia20bwa in optical (using TNG/DOLORES) and NIR (using TNG/NICS).}
\label{fig:lines_bwa}
\end{figure*}

Several lines were detected in the optical spectrum of Gaia20bwa measured with the TNG, including lines from the \ion{H}{i} Balmer series from H$\alpha$ to H$\delta$, the \ion{Na}{i} D line, the \ion{O}{i} triplet at 7771/4/5 \AA, the \ion{O}{i} line at 8446 \AA, and three lines of \ion{Ca}{ii} (Fig. \ref{fig:lines_bwa}). The \ion{H}{i} Balmer series from H$\alpha$ to H$\gamma$ were also detected in the spectrum taken with the LT. The Br$\gamma$ and Pa$\beta$ lines were detected in the NIR spectrum obtained with the TNG.

\begin{figure*}
\centering
\includegraphics[width=\textwidth]{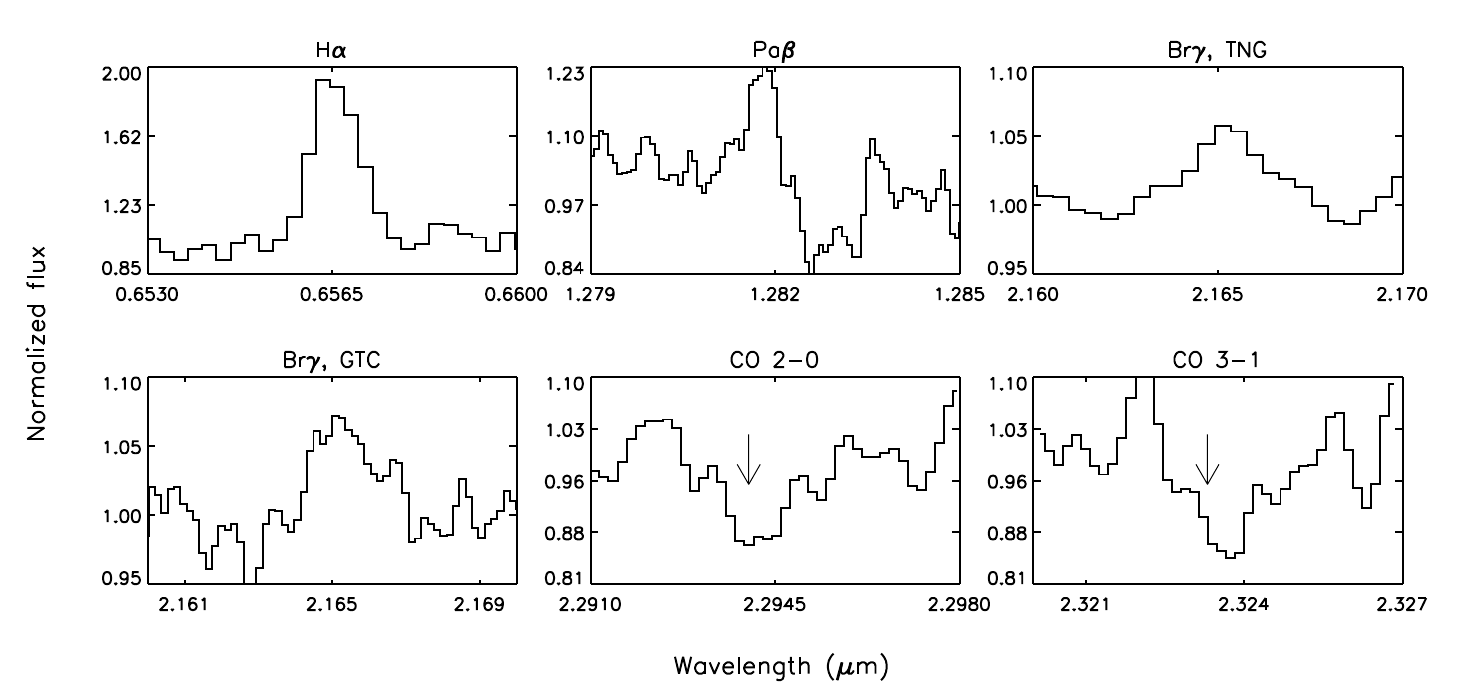}
\caption{Lines detected toward Gaia20fgx (using TNG/DOLORES and NICS and GTC/EMIR).}
\label{fig:lines_fgx}
\end{figure*}

Due to the lower S/N of the fainter Gaia20fgx, only the H$\alpha$ line of the \ion{H}{i} Balmer series was detected in the optical spectra measured with both the TNG and LT.
The Br$\gamma$ line was detected in the NIR spectrum observed with the TNG. A few additional lines were also detected in the medium resolution spectrum obtained with the GTC during the fading of the source: the CO 2-0 and 3-1 bandhead features in absorption, and the Pa$\beta$ line detected as an inverse P Cygni profile (Fig. \ref{fig:lines_fgx}).
The line parameters obtained from Gaussian fitting are shown in Table \ref{tab:lines}.

\begin{figure}
\centering
\includegraphics[width=9cm]{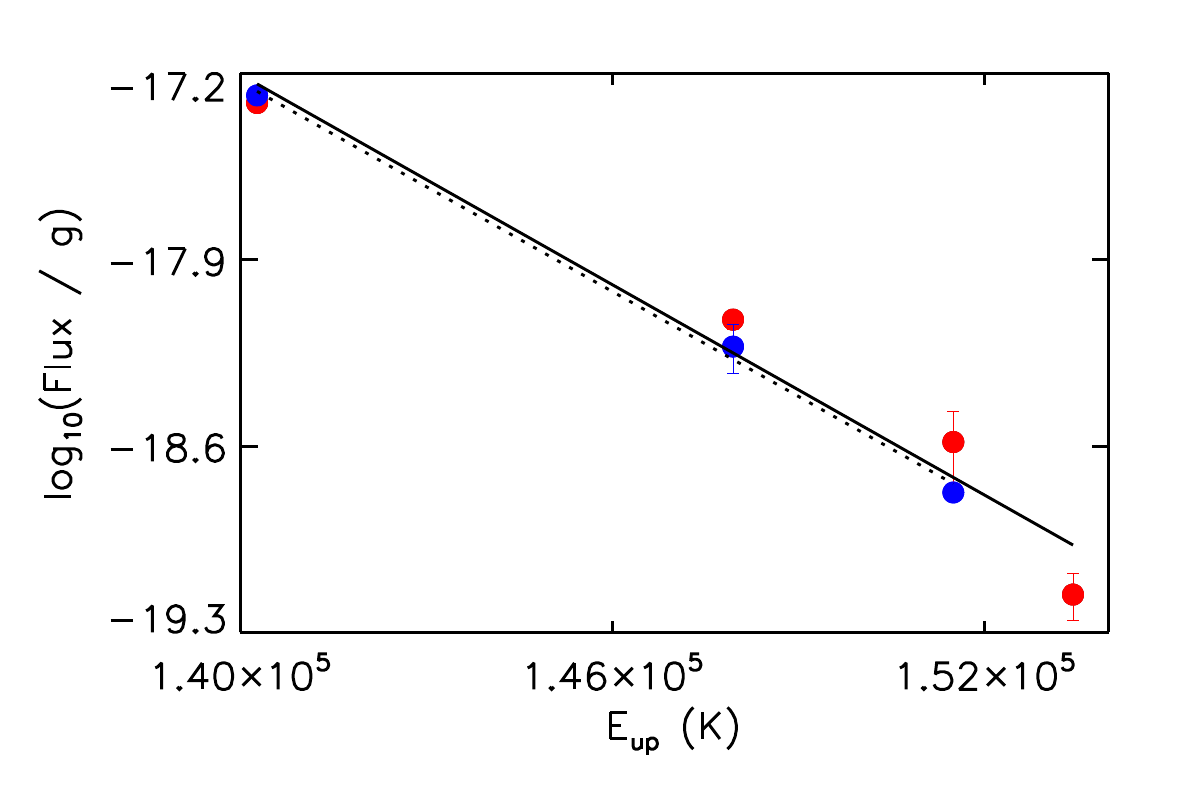}
\caption{Excitation diagram for the hydrogen Balmer lines of Gaia20bwa based on data from the TNG (red dots) and the Liverpool Telescope (blue dots). Some of the error bars are smaller than the symbol sizes. The solid line is the linear fit to the TNG data, while the dashed line is the fit to the LT data. Both lines result in a temperature of $\sim$7600~K.
}
\label{fig:ex_20bwa}
\end{figure}

Detecting several lines of the Balmer series for Gaia20bwa allows us to determine the excitation temperature of the gas that emits these lines. 
In Fig. \ref{fig:ex_20bwa} we plotted the line fluxes divided by the statistical weights of the energy levels in logarithmic scale as a function of the energy of the levels and applied a linear fit to the data points. The inverse of the slope of the lines gave an excitation temperature of $\sim$7600~K for both the TNG and LT data. This value is higher than the photospheric temperature of CTTS, as well as the 3142.5$\pm$16.0~K effective temperature derived for Gaia20bwa \citep{DaRio2016}, indicating that the detected hydrogen lines originate from gas participating in the accretion process, located close to the magnetospheric hotspot, or residing in the accretion column.
However, the $\sim$7600~K temperature derived above is an upper limit, given that Balmer lines emitted in the accretion columns are likely to be optically thick.

We can use the $\sim$7600~K temperature derived above to interpret the trend seen in the r versus [r--i] color--magnitude diagram (Fig. \ref{fig:colormag_wise}). Large amplitude optical brightness variations in CTTS in the optical regime can be produced by variations of photospheric hot spots, resulting from variable accretion rates \citep{Carpenter2001,Scholz2009}. We applied the method described by \citet{Scholz2009} to model the brightness and color variations of Gaia20bwa during the fading. We found that the amplitude and slope of the color--magnitude diagram can be reproduced by cooling of a hot spot covering some 1.5\% of the stellar surface from a temperature of $\sim$7600 K to the photospheric temperature.

\begin{table*}
\centering
\caption{Parameters of the lines detected toward Gaia20bwa and Gaia20fgx.}
\label{tab:lines}
%\resizebox{9cm}{!}{
\begin{tabular}{cccccc}
\hline
Line&   Telescope&  Center&   EW& FWHM&  $f_\mathrm{line}$\\
    &            &   (\AA)&        (\AA)& (\AA)& (W/m$^2$)\\
\hline
\multicolumn{6}{c}{Gaia20bwa}\\
\hline

H$\alpha$& TNG& 6567.3$\pm$0.3& $-$153.6$\pm$10& 12.3$\pm$0.4& (8.8$\pm${1.3})$\times$10$^{-17}$\\

H$\alpha$& LT& 6567.8$\pm$0.1& $-$152$\pm$6& 13.8$\pm$0.1& 
(9.4$\pm${1.4})$\times$10$^{-17}$\\

H$\beta$& TNG& 4863.3$\pm$1.5& $-$58$\pm$5& 12.6$\pm$1.0& 
(2.4$\pm${0.6})$\times$10$^{-17}$\\

H$\beta$& LT& 4866.1$\pm$0.1& $-$108$\pm$7& 14.5$\pm$0.1& 
(1.9$\pm${0.4})$\times$10$^{-17}$\\

H$\gamma$& TNG& 4342.4$\pm$2.4& $-$46.4$\pm$9.5& 16.0$\pm$2.0& (1.3$\pm${0.7})$\times$10$^{-17}$\\

H$\gamma$& LT& 4341.1$\pm$0.3& $-$31.8$\pm$2& 15.2$\pm$0.1& 
(8.4$\pm${2.5})$\times$10$^{-18}$\\

H$\delta$& TNG& 4103.1$\pm$0.8& $-$19.2$\pm$5& 6.5$\pm$1.9& 
(5.0$\pm${2.5})$\times$10$^{-18}$\\

\ion{Ca}{ii}& TNG&  8502.3$\pm$0.1& $-$31.3$\pm$2.9& 10.9$\pm$1.3& (4.4$\pm${0.9})$\times$10$^{-17}$\\

\ion{Ca}{ii}& TNG& 8546.6$\pm$0.3& $-$30.8$\pm$2& 11.4$\pm$1.0& (4.3$\pm${0.9})$\times$10$^{-17}$\\

\ion{Ca}{ii}& TNG& 8666.2$\pm$0.6& $-$29.2$\pm$1.3& 11.7$\pm$0.8& (4.4$\pm${0.9})$\times$10$^{-17}$\\

\ion{O}{i}& TNG& 7775.3$\pm$0.3& $-$0.9$\pm$0.1& 6.2$\pm$0.5& (1.0$\pm${0.5})$\times$10$^{-18}$\\

\ion{O}{i}& TNG& 8449.0$\pm$0.7& $-$2.7$\pm$0.8& 7.6$\pm$1.6&  (3.5$\pm${1.4})$\times$10$^{-18}$\\

\ion{Na}{i} D&  TNG& 5893.3$\pm$0.4& $-$5.7$\pm$0.6& 6.2$\pm$0.5& (3.8$\pm${1.9})$\times$10$^{-18}$\\

\ion{Na}{i} D&  TNG& 5900.5$\pm$0.1& $-$4.9$\pm$0.3& 4.3$\pm$0.1& (4.1$\pm${2.1})$\times$10$^{-18}$\\

Pa $\beta$& TNG& 12812$\pm$1& $-$8.5$\pm$0.5& 13.5$\pm$0.5& 
(1.4$\pm${0.5})$\times$10$^{-17}$\\

Br $\gamma$& TNG& 21647$\pm$2& $-$13.7$\pm$0.3& 23$\pm$1& 
(8.8$\pm${3.1})$\times$10$^{-18}$\\

\hline
\multicolumn{6}{c}{Gaia20fgx}\\
\hline 

H$\alpha$& TNG& 6564.8$\pm$2& $-$12.3$\pm$1& 10$\pm$2& (4.0$\pm${0.8})$\times$10$^{-18}$\\

Br $\gamma$& TNG& 21683$\pm$4& $-$10$\pm$1& 48$\pm$6& (4.0$\pm${2.0})$\times$10$^{-18}$\\

Br $\gamma$& GTC& 21652$\pm$1& $-$1.4$\pm$0.2& 20$\pm$3& (7.0$\pm${2.1})$\times$10$^{-19}$\\

\hline
\end{tabular}
%}
\end{table*}

\subsection{SEDs and stellar parameters}
\label{sect:sed}

\begin{figure}
\centering
\includegraphics[width=8cm]{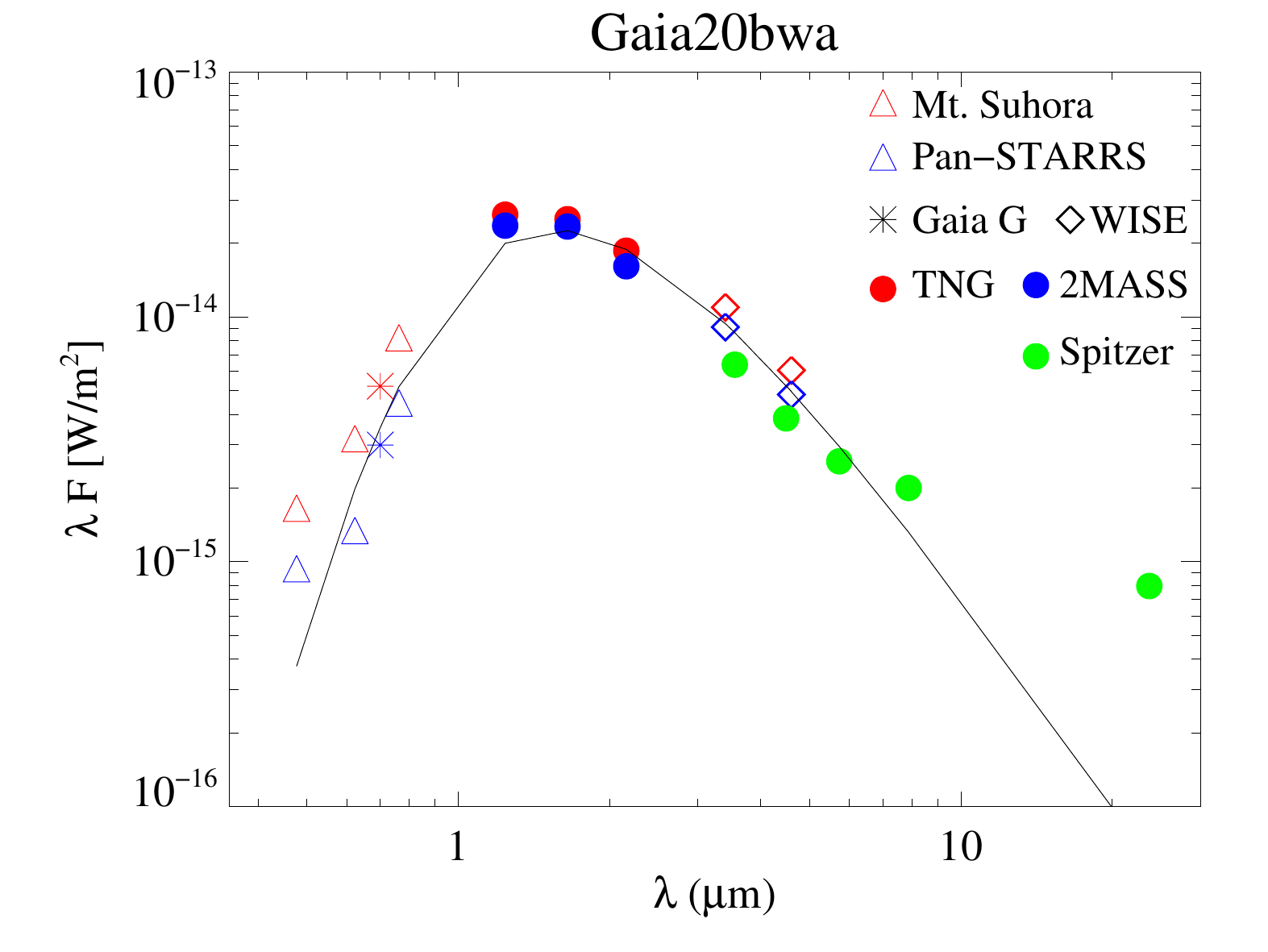}
\includegraphics[width=8cm]{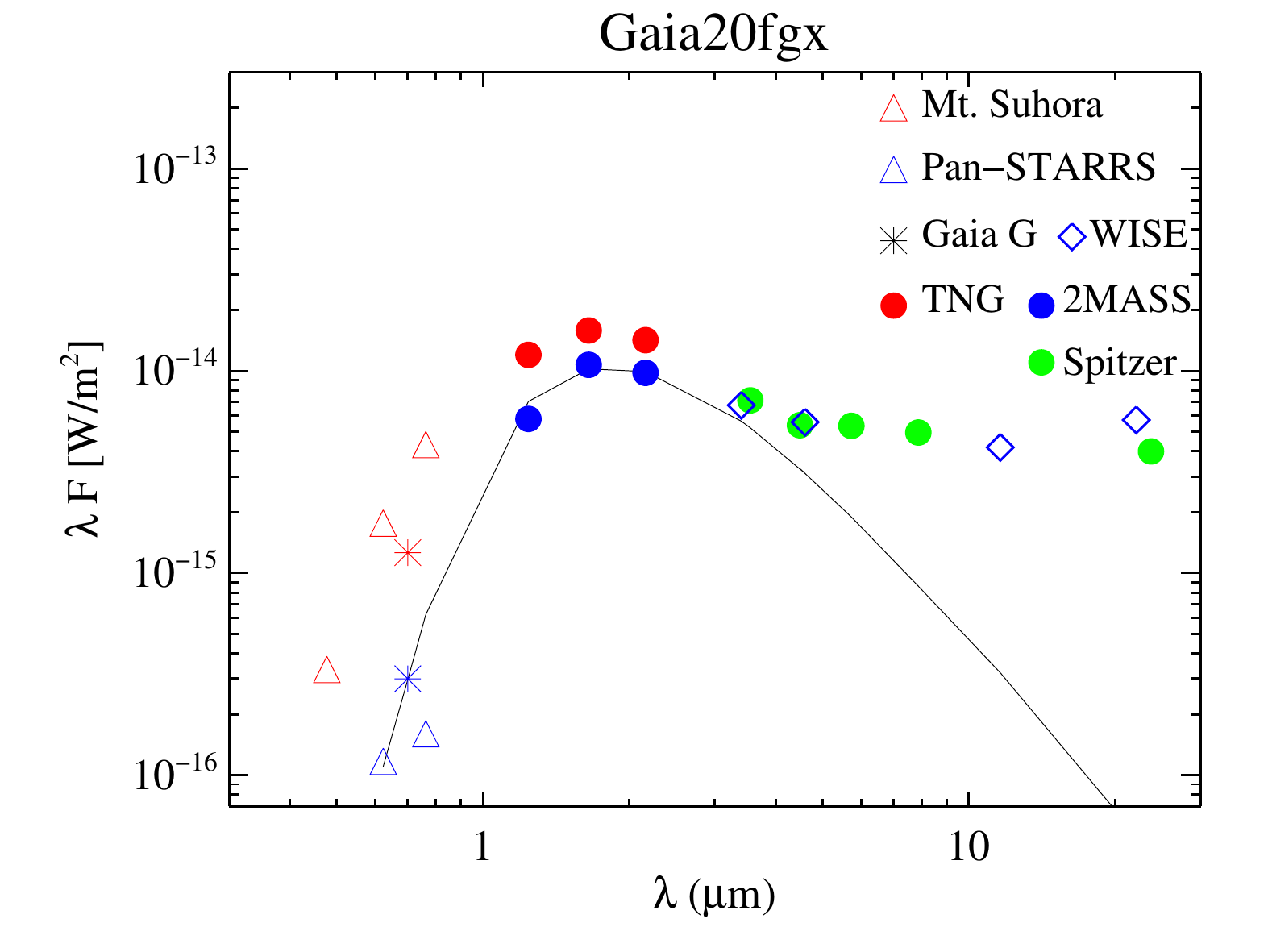}
\caption{SEDs of Gaia20bwa and Gaia20fgx in the bright (red symbols) and faint (blue and green symbols) states.
The overplotted black body curves correspond to a temperature of 3300~K and an $A_V$ of 3 mag for Gaia20bwa, and a temperature of 3700~K and an $A_V$ of 7 mag for Gaia20fgx.
}
\label{fig:sed}
\end{figure}

The SEDs of Gaia20bwa and Gaia20fgx are shown in Fig. \ref{fig:sed}. The data points for the bright state (red symbols) are based on our follow-up observations with the TNG and Mt. Suhora telescopes, the \textit{Gaia} $G$-band data measured close to the TNG and Mt. Suhora observations, and on the last published \textit{WISE} W1 and W2 fluxes. The data points for the faint state are based on archival data from Pan-STARRS, \textit{Gaia}, \textit{WISE}, 2MASS and \textit{Spitzer}. 
We computed black body functions for a range of temperatures and visual extinctions and compared them to the SEDs by visual inspection. We used the temperature and $A_V$ values estimated from the SEDs to derive stellar parameters for the sources. We explain the results for each source below. 

\subsubsection{Gaia20fgx}

The SED of Gaia20fgx in the faint state is consistent with a temperature of 3700$\pm$300 K and a visual extinction of $A_V=7.0 \pm 0.7$~mag, which is slightly above the value that was derived for the faint state based on the $J-H$ versus $H-K_s$ plot, and is consistent with the $\sim$7.3 mag value derived by \citet{Chen2020}. Similarly, the SED in the bright state is more consistent with a higher $A_V$ compared to the value derived from the $J-H$ versus $H-K_s$ plot, with an $A_V$ of $5\pm0.5$ mag.
The 3700 K temperature is between the M0 and M1 spectral types \citep{PecautMamajek2013}. 
The luminosity of Gaia20fgx can be derived from the observed magnitudes corrected for extinction and assuming a bolometric correction (BC). 
We used the bolometric correction value corresponding to the average of the values for the M0 and M1 spectral types in the $J$ band (BC$_J$) of 5-30 Myr stars from Table 6 of \citet{PecautMamajek2013}. The luminosity of Gaia20fgx is then
\begin{equation}
\label{luminosity}
{\rm{log}} \left( \frac{L_\star}{L_\odot} \right) = 0.4 ( M_{{\rm{bol}},\odot} - M_{\rm{bol}} ), 
\end{equation}
where $M_{{\rm{bol}},\odot}$ is the bolometric magnitude of the Sun \citep{Mamajek2015}, and the bolometric magnitude of the source is $M_{\rm{bol}} = m_J - 5 {\rm{log}}(d/10[pc]) + {\rm{BC}}_J$ where $m_J$ is the extinction corrected $J$ magnitude of 12.57$\pm$0.03 mag (assuming an $A_V=7$ mag). This results in $L_\star = 1.02\pm0.20~L_\odot$ for Gaia20fgx. 
Based on the $L_\star$ and $T_{\rm{eff}}$ derived above and the evolutionary tracks by \citet{Siess2000}, we determined a stellar mass of $M_\star=0.53\pm0.10~M_\odot$.
Assuming that the central object emits as a black-body, the stellar radius can be derived as
\begin{equation}
\label{radius}
R_\star = \frac{1}{2 T^2_{\rm{eff}}} \sqrt{\frac{L_\star}{\pi \sigma}}
\end{equation}
where $\sigma$ is the Stefan-Boltzmann constant. The $R_\star$ for Gaia20fgx is 2.46$\pm$0.52 $R_\odot$ using this method.
The stellar parameters derived for Gaia20fgx in the faint and bright states are summarized in Table \ref{tab:stellar_param}.

\subsubsection{Gaia20bwa}

The SED of Gaia20bwa in the faint state is consistent with a temperature of 3300$\pm$200 K and a visual extinction of $A_V=3.0 \pm 0.5$~mag. This temperature is consistent with the 3142.5$\pm$16.04~K derived by \citet{DaRio2016}, but the derived $A_V$ value is above the 0.9$\pm$0.35 mag suggested by \citet{DaRio2016}. These parameters (3300~K temperature and 3.0 mag visual extinction) are also consistent with SED in the bright state.
A temperature of 3300~K is close to the M3 spectral type \citep{PecautMamajek2013}. We used the bolometric correction value corresponding to the M3 spectral type in the $J$ band (BC$_J$) of 5-30 Myr stars from Table 6 of \citet{PecautMamajek2013}. The extinction corrected $J$ magnitude is 12.03$\pm$0.03 mag (assuming an $A_V=3.0$ mag). Using Eqn. \ref{luminosity}, we derived a luminosity of $0.38\pm0.06$ $L_\odot$. Based on the $L_\star$ and $T_{\rm{eff}}$ derived above and the evolutionary tracks by \citet{Siess2000}, we determined a stellar mass of $M_\star=0.28\pm0.06~M_\odot$. Using Eqn. \ref{radius}, we derived a radius of 1.89$\pm$0.32 $R_\odot$.
The stellar parameters derived for Gaia20bwa are summarized in Table \ref{tab:stellar_param}. 
The temperature, $A_V$, $L_\star$, and $M_\star$ derived for Gaia20bwa by \citet{DaRio2016} are also listed in Table \ref{tab:stellar_param}. The $R_\star$ corresponding to these parameters was derived using Eqn. \ref{radius} based on the temperature and luminosity from \citet{DaRio2016}.
Although the temperature derived here for Gaia20bwa is consistent with the value derived by \citet{DaRio2016} within the uncertainties, the stellar luminosity, mass, and radius derived here are about factors of 2, 1.4, and 1.3 larger, respectively. The difference between the stellar parameters derived here and by \citet{DaRio2016} is mostly due to the higher visual extinction we found from the SED fit.

\begin{table*}
\centering
\caption{Summary of the stellar parameters of Gaia20bwa and Gaia20fgx.}
\label{tab:stellar_param}
\begin{tabular}{rccccc}
\hline
& $T$ (K)& $A_V$ (mag)& $L_\star$ ($L_\odot$)& $M_\star$ ($M_\odot$)& $R_\star$ ($R_\odot$)\\
\hline

Gaia20bwa faint \& bright states&     
3300$\pm$200& 3.0$\pm$0.5& 0.38$\pm$0.06& 0.28$\pm$0.06& 1.89$\pm$0.32\\

Gaia20bwa based on \citet{DaRio2016}& 
3142.5$\pm$16.04& 0.9$\pm$0.35& 0.20$\pm$0.02& 0.206$\pm$0.008& 1.51$\pm$0.15\\

\hline

Gaia20fgx faint state& 
3700$\pm$300& 7.0$\pm$0.7& 1.02$\pm$0.20& 0.53$\pm$0.10& 2.46$\pm$0.52\\

Gaia20fgx bright state& 
3700$\pm$300& 5.0$\pm$0.5& 0.61$\pm$0.12& 0.47$\pm$0.10& 1.90$\pm$0.40\\

\hline
\end{tabular}
\end{table*}

\subsection{Accretion rates}

The line fluxes shown in Table 2 can be converted to line luminosities as $L_{\rm{line}}=4 \pi d^2 f_{\rm{line}}$, where $d$ is the distance of the sources, and $f_{\rm{line}}$ is the extinction-corrected flux of the lines.
For the accretion parameters derived in this Section, we use the visual extinctions estimated from the SEDs when available, rather than the values derived from the $J-H$ versus $H-K_s$ plot. The SEDs allow to better constrain the visual extinctions than three data points (the $J$, $H$, and $K_s$ magnitudes). 
The accretion parameters derived from the different lines also confirm this, as they are more consistent when derived using the visual extinctions estimated from the SED, rather than from the $J-H$ versus $H-K_s$ plot. 
The visual extinctions derived from the SEDs in the bright state are 3 mag for Gaia20bwa and 5 mag for Gaia20fgx, while in the faint state they are 3 mag and 7 mag, respectively.
For the estimate based on the Br$\gamma$ line observed with the GTC, we used the extinction of $\sim$4.1 mag derived from the $J-H$ versus $H-K_s$ plot.
However, we found, that the SEDs suggest 20-30\% higher visual extinctions compared to the $J-H$ versus $H-K_s$ plot. Therefore, to be consistent with the $A_V$ values derived from the SEDs, we used a value of 5.5 mag for the GTC data, a 25\% higher value than derived from the $J-H$ versus $H-K_s$ plot.
For the extinction correction, we assumed an $R_V$ of 5.5 for Gaia20bwa, as suggested by \citet{DaRio2016} as an average value for the Orion A cloud, while we assumed an $R_V$ of 3.1 for Gaia20fgx. However, this does not make a significant difference in the derived accretion parameters, and affects the accretion parameters by $\sim$10\% or less.
For the distance of Gaia20bwa we adopted 410$^{+12}_{-11}$ pc \citep{BailerJones2021}, while for the distance of Gaia20fgx we assumed the $\sim$816$\pm$53 pc derived in Sec. \ref{sect_distance} for the distance of Cep OB3.  
We derived the accretion luminosities from the line luminosities based on the relations provided by \citet{Alcala2017}.  
The accretion luminosities can then be converted to accretion rates using the formula
\begin{equation}
\nonumber
\dot{M}_{\rm{acc}} = 1.25 \frac{L_{\rm{acc}} R_\star}{G M_\star} 
\end{equation}
for which an inner-disk radius of 5 $R_\star$ was assumed \citep{Hartmann1998}. 
For Gaia20bwa, we used the radius of 1.89$\pm$0.32 $R_\odot$ and stellar mass of $0.28\pm0.06~M_\odot$ derived above to provide one estimate of the accretion rates, and the radius of 1.51$\pm$0.15 $R_\odot$ and stellar mass of 0.206$\pm0.008~M_\odot$ for another estimate \citep{DaRio2016}.
For Gaia20fgx, we adopt the $R_\star$=2.46$\pm$0.52~$R_\odot$ and $M_\star=0.53\pm0.1$~$M_\odot$ values derived above for the faint state, and $R_\star$=1.90$\pm$0.40~$R_\odot$ and $M_\star=0.47\pm0.1$~$M_\odot$ values derived above for the bright state.

The accretion luminosities and rates are shown in Table 2 for both sets of stellar parameters from Table \ref{tab:stellar_param} for both sources.
For the stellar parameters derived above for Gaia20bwa, the accretion luminosities are in the range between $(6.1 \pm 2.4) \times 10^{-2}$ $L_\odot$ and $5.78 \pm 3.11 \times 10^{-1}$ $L_\odot$, and the accretion rates are in the range between $(1.65 \pm 0.79) \times 10^{-8}$ $M_\odot$ yr$^{-1}$ and $(1.60 \pm 0.9) \times 10^{-7}$ $M_\odot$ yr$^{-1}$. Based on the stellar parameters from \citet{DaRio2016}, the accretion luminosities are in the range between $(1.64\pm0.85) \times 10^{-2}$ $L_\odot$ and $(1.67\pm4.35) \times 10^{-1}$ $L_\odot$, and the accretion rates are in the range between $(4.82 \pm 2.55) \times 10^{-9}$ $M_\odot$ yr$^{-1}$ and $(4.89 \pm 13.78) \times 10^{-8}$ $M_\odot$ yr$^{-1}$.

The accretion luminosities for Gaia20fgx during its bright state based on the TNG data are $(0.95 - 5.21) \times 10^{-1}$ $L_\odot$ and $(2.77 - 3.58) \times 10^{-1}$ $L_\odot$ based on the H$\alpha$ and the Br$\gamma$ lines, respectively, taking into account both sets of stellar parameters (Table \ref{tab:acc_parameters}).
The accretion rates during the bright state are $(1.54 - 9.68) \times 10^{-8}$ $M_\odot$ yr$^{-1}$ and $(4.48 - 6.64) \times 10^{-8}$ $M_\odot$ yr$^{-1}$ based on the H$\alpha$ and the Br$\gamma$ lines, respectively. 
The accretion luminosity and rate derived for Gaia20fgx based on the Br$\gamma$ line measured with the GTC about half a year after the TNG measurements are almost a factor of 10 below those derived from the TNG data during the bright state. Since the Pa$\beta$ line detected with the GTC also has an absorption component, we did not use it to derive the accretion luminosity and rate.

\begin{table*}
\centering
\caption{Accretion luminosities and rates for Gaia20bwa and Gaia20fgx. 
The accretion luminosities calculated from the fluxes based on \citet{Alcala2017}. For Gaia20bwa, $L_{\rm{acc,1}}$ and $\dot{M}_{\rm{acc,1}}$ correspond to the $M_\star$ and $R_\star$ values derived for the faint and bright states, while $L_{\rm{acc,2}}$ and $\dot{M}_{\rm{acc,2}}$ correspond to the $M_\star$ and $R_\star$ values derived by \citet{DaRio2016}. For Gaia20fgx, $L_{\rm{acc,1}}$ and $\dot{M}_{\rm{acc,1}}$ correspond to the $M_\star$ and $R_\star$ values derived for the faint state, and  $L_{\rm{acc,2}}$ and $\dot{M}_{\rm{acc,2}}$ correspond to the $M_\star$ and $R_\star$ values derived for the bright state.
}
\label{tab:acc_parameters}
\begin{tabular}{cccccc}
\hline
Line&   Telescope&  $L_{\rm{acc,1}}$& $\dot{M}_{\rm{acc,1}}$& $L_{\rm{acc,2}}$& $\dot{M}_{\rm{acc,2}}$\\
& & ($10^{-1} L_\odot$)& ($10^{-8} M_\odot$ yr$^{-1}$)& ($10^{-2} L_\odot$)& ($10^{-9} M_\odot$ yr$^{-1}$)\\
\hline
\multicolumn{6}{c}{Gaia20bwa}\\
\hline

H$\alpha$& TNG& 
1.36$\pm$0.30& 3.68$\pm$1.29& 2.09$\pm$0.46& 6.14$\pm$1.50\\

H$\alpha$& LT&
1.47$\pm$0.32& 3.97$\pm$1.39& 2.25$\pm$0.49& 6.61$\pm$1.60\\

H$\beta$& TNG&
4.46$\pm$1.26& 12.04$\pm$4.73& 3.95$\pm$1.11& 11.58$\pm$3.48\\

H$\beta$& LT& 
3.41$\pm$0.84& 9.23$\pm$3.40& 3.02$\pm$0.75& 8.87$\pm$2.40\\

H$\gamma$& TNG& 
4.54$\pm$2.51& 12.26$\pm$7.56& 3.50$\pm$1.94& 10.26$\pm$5.79\\

H$\gamma$& LT& 
2.79$\pm$0.90& 7.55$\pm$3.19& 2.15$\pm$0.70& 6.32$\pm$2.17\\

H$\delta$& TNG& 
2.13$\pm$1.10& 5.74$\pm$3.35& 1.64$\pm$0.85& 4.82$\pm$2.55\\

\ion{Ca}{ii}& TNG&
5.46$\pm$1.43& 14.76$\pm$5.59& 16.66$\pm$4.35& 48.87$\pm$13.78\\

\ion{Ca}{ii}& TNG& 
4.07$\pm$1.10& 11.00$\pm$4.23& 12.84$\pm$3.48& 37.66$\pm$10.97\\

\ion{Ca}{ii}& TNG& 
3.90$\pm$1.07& 10.53$\pm$4.08& 13.19$\pm$3.62& 38.71$\pm$11.40\\

\ion{O}{i}& TNG& 
1.07$\pm$0.57& 2.90$\pm$1.74& 1.90$\pm$1.00& 5.56$\pm$2.99\\

\ion{O}{i}& TNG& 
1.44$\pm$0.67& 3.88$\pm$2.09& 3.89$\pm$1.81& 11.40$\pm$5.44\\

\ion{Na}{i} D&  TNG& 
3.46$\pm$1.83& 9.35$\pm$5.57& 5.46$\pm$2.88& 16.02$\pm$8.62\\

\ion{Na}{i} D&  TNG& 
5.78$\pm$3.11& 15.60$\pm$9.41& 9.13$\pm$4.92& 26.78$\pm$14.71\\

Pa $\beta$& TNG& 
0.61$\pm$0.24& 1.65$\pm$0.79& 3.17$\pm$1.26& 9.31$\pm$3.83\\

Br $\gamma$& TNG& 
1.14$\pm$0.46& 3.09$\pm$1.51& 8.33$\pm$3.32& 24.44$\pm$10.08\\

\hline
\multicolumn{6}{c}{Gaia20fgx}\\
\hline 

H$\alpha$& TNG& 
5.21$\pm$1.40& 9.68$\pm$3.78& 9.50$\pm$2.55& 15.37$\pm$6.18\\

Br $\gamma$& TNG& 
3.58$\pm$1.93& 6.64$\pm$4.04& 27.70$\pm$14.96& 44.83$\pm$27.68\\

Br $\gamma$& GTC& 
0.37$\pm$0.13& 0.70$\pm$0.32& 3.71$\pm$1.35& 6.01$\pm$2.83\\

\hline
\end{tabular}
\end{table*}

\section{Discussion}
\label{sect_discussion}

\subsection{The origin of the brightening events}

Both targets show a brightening on a time-scale, which is typical of EXors (\citealp{Herbig1989,Herbig2008}). 
However, the amplitude of the brightening for Gaia20bwa ($\sim$0.5 mag) is lower than expected for EXors, while the 2.5 mag brightening for Gaia20fgx is closer to that expected for EXors. In addition to the time-scale and amplitude of the brightening event reported as a \textit{Gaia} alert for Gaia20fgx, it showed a brightening event earlier, between early 2018 and early 2019 (Fig. \ref{fig:light}). Based on the currently available data, there is no evidence that the brightening of Gaia20bwa is also a recurring event. To understand the brightening events of the two targets, we compare their accretion parameters to those derived for other young stars.

\begin{figure}
\centering
\includegraphics[width=9cm, trim=0 0 0 1.2cm,clip]{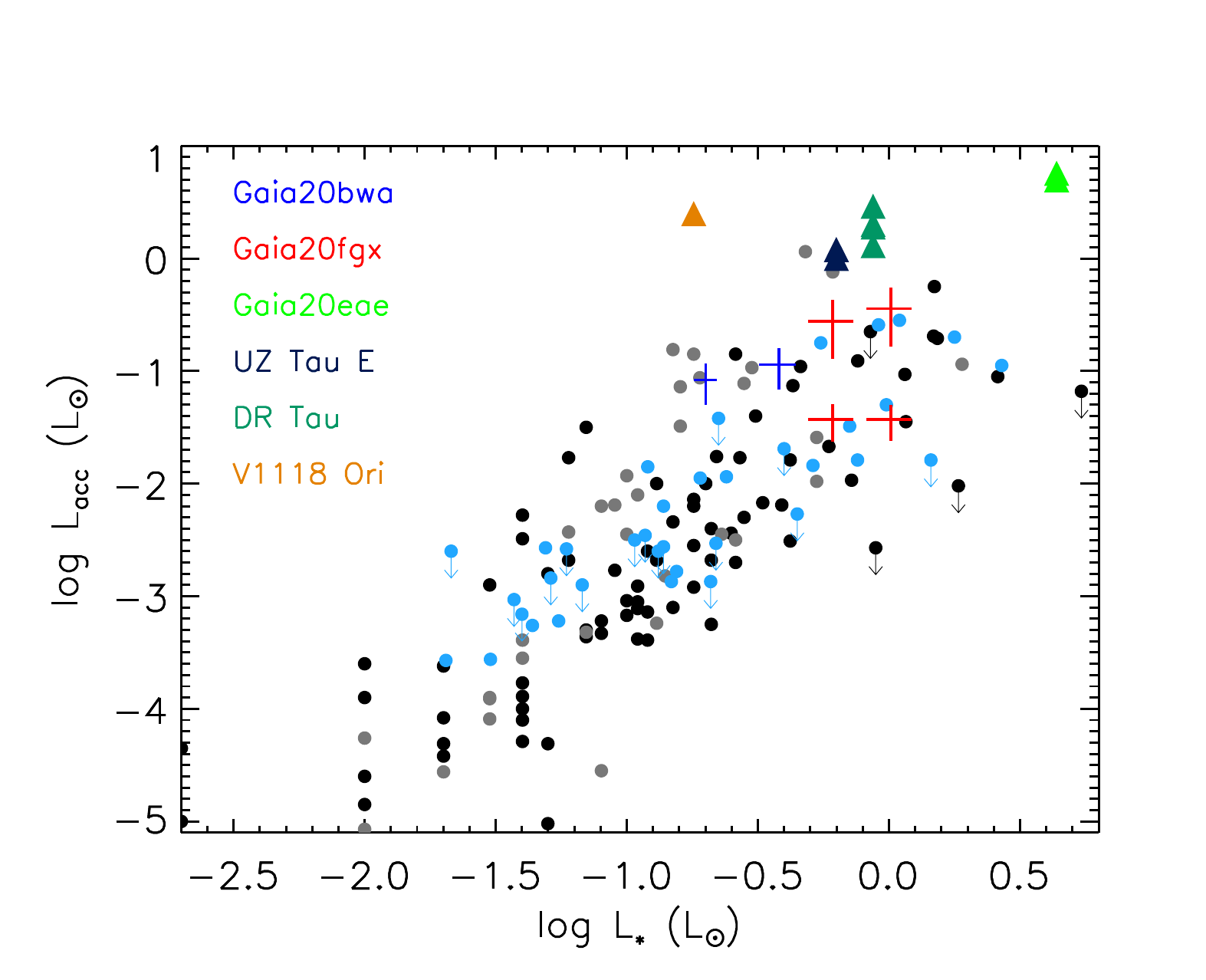}
\includegraphics[width=9cm, trim=0 0 0 1.2cm,clip]{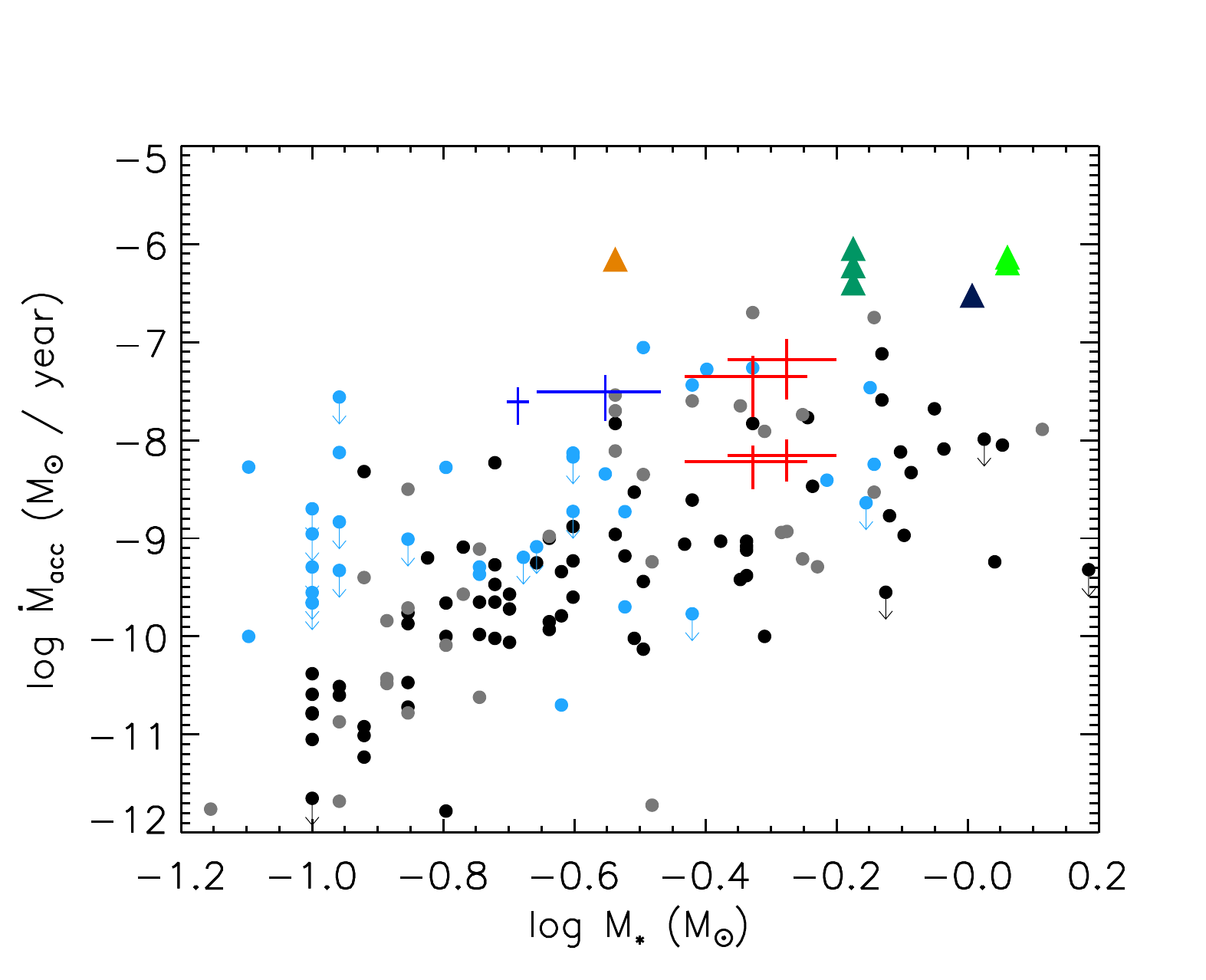}
\caption{Accretion luminosities versus stellar luminosities (top panel), as well as accretion rates versus stellar masses (bottom panel) and their comparison to confirmed EXors (triangles) and samples of CTTS in the Lupus (black symbols, \citealp{Alcala2019}), Chamaeleon I (grey symbols, \citealp{Manara2019}), and NGC 1333 (light blue symbols, \citealp{Fiorellino2021}) regions. The upper limits are marked with downward arrows.}
\label{fig:accr_lum_rate}
\end{figure}

The accretion luminosities versus the stellar luminosities as well as the accretion rates versus the stellar masses are plotted in Fig. \ref{fig:accr_lum_rate}. Samples of CTTS are also plotted for comparison toward the Lupus (black symbols, \citealp{Alcala2019}), the Chamaeleon I (grey symbols, \citealp{Manara2019}), and the NGC 1333 (light blue symbols, \citealp{Fiorellino2021}) regions. The stellar luminosity of Gaia20bwa is close to the median value of the three samples, while its stellar mass is slightly below the median value of CTTS. Gaia20fgx is close to the most luminous end of the plotted CTTS, and its mass is above the median value of the plotted CTTS. While the accretion luminosities and rates for both Gaia20bwa and Gaia20fgx in their bright state are toward the upper end among CTTS with similar luminosities and masses, they still follow the trend seen in the accretion luminosity versus stellar luminosity and the accretion rate versus stellar mass in the CTTS samples. This is not the case for the few examples of EXors (triangles in Fig. \ref{fig:accr_lum_rate}), which are clearly above the general trends seen in the CTTS samples. One of the EXors plotted is the newly confirmed EXor corresponding to the \textit{Gaia} alerted Gaia20eae \citep{CruzSaenzdeMiera2022}. The accretion luminosities and rates of the other EXors are from \citet{Lorenzetti2009}. For UZ Tau E we used a stellar luminosity of $\sim$0.6 $L_\odot$ and a mass of $\sim$1.0 $M_\odot$ \citep{Prato2002}. For DR Tau we adopted a stellar luminosity of 0.87 $L_\odot$ \citep{Muzerolle2003}, which, together with its K7 spectral type, corresponds to a stellar mass of $\sim$0.67 $M_\odot$ \citep{Siess2000}. For V1118 Ori we adopted a stellar luminosity of 0.18 $L_\odot$ and a stellar mass of 0.29 $M_\odot$ \citep{Giannini2017}.
For a better comparison, for Gaia20bwa, Gaia20fgx, and Gaia20eae \citep{CruzSaenzdeMiera2022}, and the EXors from \citet{Lorenzetti2009} we used the accretion luminosities and rates calculated from the Br$\gamma$ line.
Based on this comparison of the accretion versus stellar parameters, not only Gaia20bwa, but also Gaia20fgx is closer to CTTS than to EXors.

Brightness variations for young stars also occur due to a change of the circumstellar extinction, not only due to a change of the accretion rate. As seen in Fig. \ref{fig:colormag_wise} in the $g$ versus [$g-r$] color--magnitude diagram of Gaia20fgx, the brightening event corresponding to the \textit{Gaia} alert suggests a change in the visual extinction, such as suggested by the $J-H$ versus $H-K_s$ plot and the SEDs. To investigate how the accretion rate changes, we observed the NIR spectra of Gaia20fgx in its fading phase, about half a year after the first spectrum was taken in the bright state. The accretion rate derived from the Br$\gamma$ line during the fading phase ($(6.01 - 7.0) \times 10^{-9}$ $M_\odot$ yr$^{-1}$) is about a factor of 10 lower than during the bright state ($(4.48 - 6.64) \times 10^{-8}$ $M_\odot$ yr$^{-1}$). This suggests, that while the visual extinction changed between the two epochs, the brightening event of Gaia20fgx corresponding to the \textit{Gaia} alert was indeed due to an increase of the accretion rate, even if the increased accretion rate remained below the values expected for eruptive young stars.
For Gaia20bwa, while there is only one estimate of its accretion rate, the color--color diagram in Fig. \ref{fig:infra_color} indicates, that the visual extinction did not change significantly during the faint and the bright state. In addition to this, the brightening event was also seen in the WISE W1 and W2 bands.
Therefore, the brightening event for Gaia20bwa is likely also related to an increase of the accretion rate.

It was observed for several young eruptive stars, including Gaia17bpi \citep{Hillenbrand2018} and Gaia18dvy \citep{SzegediElek2020}, that their brightening occurs earlier at IR than at optical wavelengths. This phenomenon was found to be consistent with disk models that predict instabilities in the inner 0.5–1 au of accretion disks \citep{Hillenbrand2018}. Based on the \textit{Gaia} and \textit{WISE} light curves, there is no evidence of such behavior for either Gaia20bwa or Gaia20fgx, which supports the finding, that their brightenings are not related to instabilities like in the case of young eruptive stars.

\subsection{The time-scale and amplitude of the brightening events}

As discussed above from the comparison of the accretion rates, the sources studied here are most likely CTTS, rather than EXors. However, the brightness variations of CTTS typically occur on shorter time-scales than the brightenings of Gaia20bwa and Gaia20fgx, and with lower amplitudes \citep{HillenbrandFindeisen2015}. 
Though the brightening events with 0.5$-$2.5 mag amplitudes on a time-scale of a year or above are not typical of non-eruptive YSOs, a number of sources with variability on a similar time-scale were identified at NIR wavelengths \citep{ContrerasPena2017}. The sample of \citet{ContrerasPena2017} includes targets with $\Delta K_S > 1$ mag, out of which, several sources show long-term brightening, on a similar time-scale to Gaia20bwa and Gaia20fgx. However, given that the survey of \citet{ContrerasPena2017} was in the NIR, it cannot be directly compared to the \textit{Gaia} light curves of Gaia20bwa and Gaia20fgx.
The long-term variability of a sample of 72 CTTS based on optical photometry was analysed by \citet{Grankin2007}, and though most of the sources in the sample showed a variability with an amplitude of $\Delta V \leq 0.4$ mag, there are a few sources with variability amplitudes up to $\Delta V \sim 2$ mag. 
Based on the long-term optical light curves of 218 CTTS, \citet{Briceno2019} found mean variability amplitudes of $\sim$0.7 mag in the V-band and $\sim$0.6 mag in the R-band, which are close to the $\sim$1 mag variability, that is seen for Gaia20fgx between 2015 and 2018.
The information on long-term brightness variations of YSOs is limited.
Databases from photometric surveys, such as the \textit{Gaia} Science Alerts, are expected to provide more information on brightenings of young stars on a time-scale of months-years.

\subsection{Spectral properties of the sources}

\begin{figure}
\centering
\includegraphics[width=9cm, trim=0.6cm 0.35cm 0 0cm,clip]{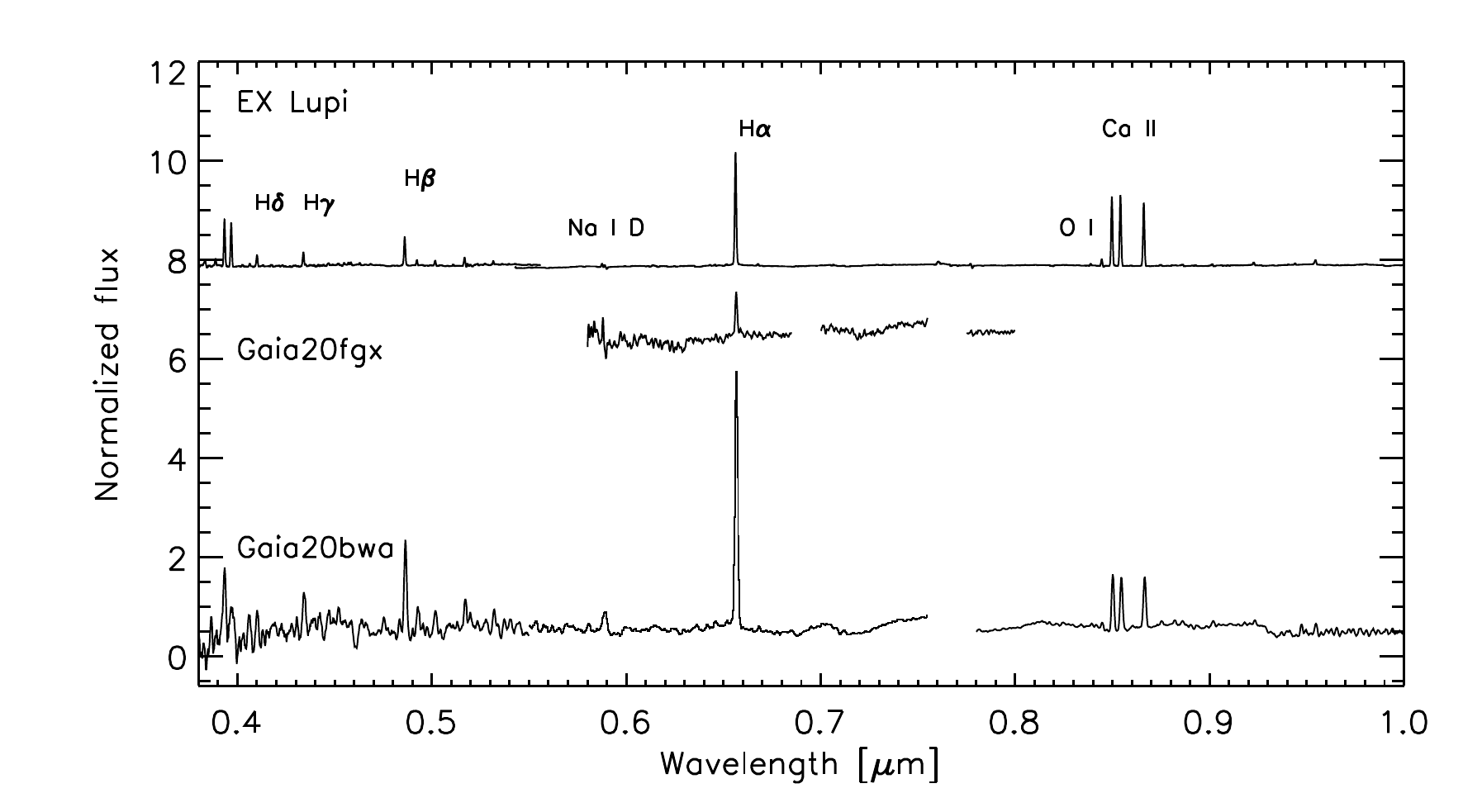}
\includegraphics[width=9cm, trim=0.6cm 0.35cm 0 0cm,clip]{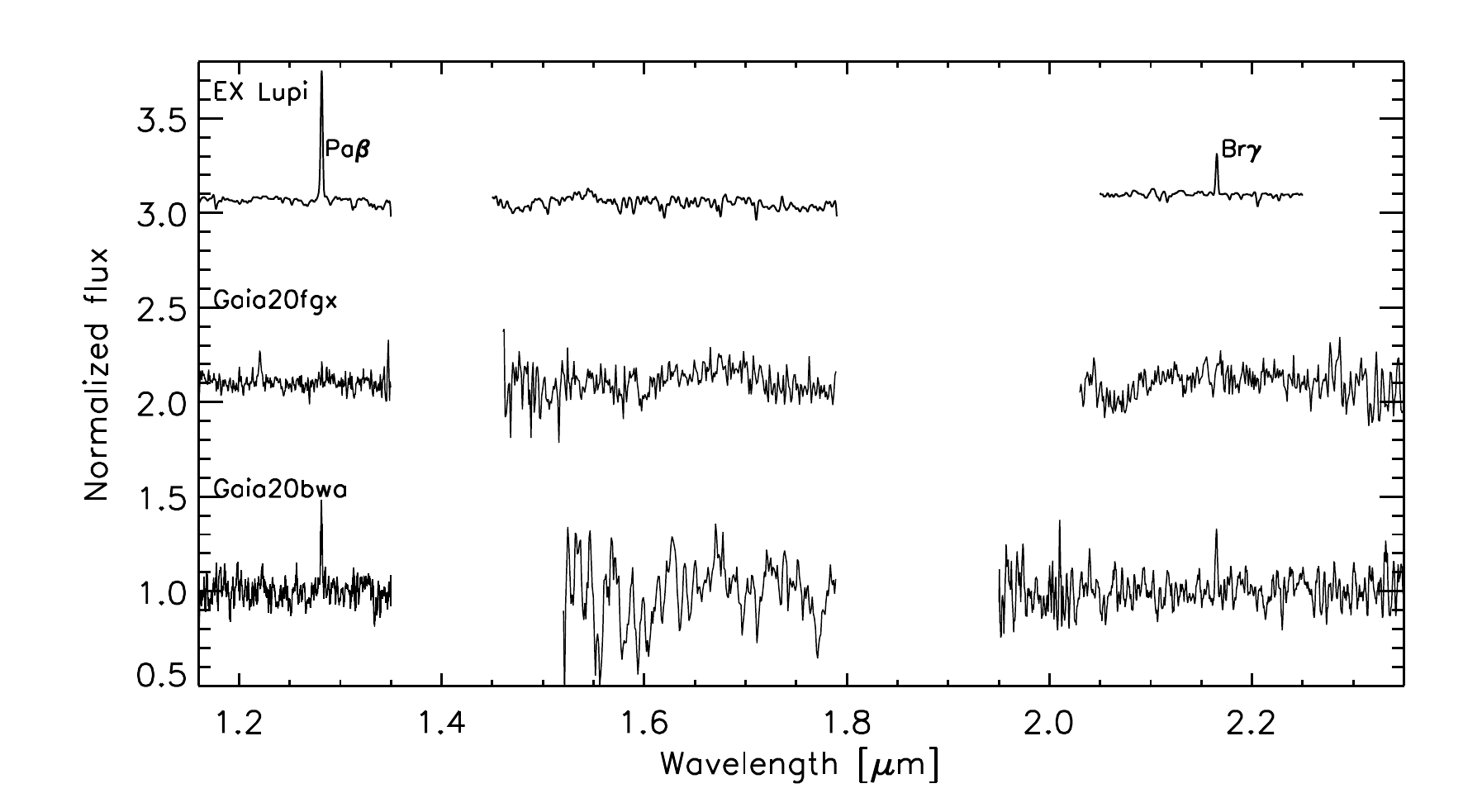}
\caption{Comparison of the optical and NIR TNG spectra of Gaia20bwa and Gaia20fgx to those of EX Lupi, the proto-typical EXor during its latest outburst (\'Abrah\'am et al., in prep.). The spectra of EX Lupi were smoothed to match their spectral resolution similar to that of the TNG spectra. For a better comparison, the optical fluxes of Gaia20bwa were multiplied by 0.6, the optical spectra of EX Lupi were multiplied by 0.4, and the NIR spectra of EX Lupi were multiplied by 2.}
\label{fig:comparison_spectra}
\end{figure}

The lines identified in the spectra of both sources are also typical of EXors (\citealp{CruzSaenzdeMiera2022} and references therein). However, the number of lines detected in the spectra is below what is typical of EXors, even when observed at low spectral resolution \citep{Lorenzetti2009}. For Gaia20fgx, the low number of detected lines in the spectra, in addition to the low spectral resolution, may also be related to its low brightness even during its bright state, when the TNG spectra were observed, and during its fading, when the GTC spectra were observed. 
As a comparison with a known EXor, the prototype EX Lupi, we plotted in Fig. \ref{fig:comparison_spectra} the TNG spectra of Gaia20bwa and Gaia20fgx together with those of EX Lupi. We used the spectra obtained during its latest, 2022 outburst \citep{Kospal2022}, as its amplitude was $\lesssim$2 mag in the $g$ band (\'Abrah\'am et al., in prep.), and as a less luminous outburst compared to the extreme eruption of EX Lupi in 2008 \citep{Abraham2009}, it provides a good comparison with our targets. Since these spectra were taken using VLT/XSHOOTER (\'Abrah\'am et al., in prep.), we smoothed them to match their resolution to our low-resolution spectra. Several spectral lines are detected in the spectra of both Gaia20bwa and EX Lupi, e.g. the Balmer lines, Pa$\beta$, Br$\gamma$. However, several lines, which were detected in the spectrum of EX Lupi, such as lines of \ion{He}{i}, Pa$\gamma$, and Pa$\delta$, are not seen in the spectra of Gaia20bwa and Gaia20fgx.
The visual extinctions and spectral types of the sources derived from the SEDs are similar to those of CTTS (e.g., \citealp{Fiorellino2021}) including EXors (\citealp{Lorenzetti2009}). 
We conclude, that the spectral properties of Gaia20bwa and Gaia20fgx, together with their accretion parameters, are more consistent with active CTTS, than with typical EXors.

\section{Summary}
\label{sect_summary}

We have analyzed the light curves and optical and NIR spectra of two young stars which had \textit{Gaia} alerts due to brightening episodes on a time-scale which is typical of the EXor class of young eruptive stars. The main results can be summarized as follows.

\begin{itemize}

\item[-] The brightening episode of Gaia20bwa occurred on a time-scale of a year and few months with an amplitude of $\sim$0.5 mag. Gaia20fgx showed two brightening episodes on a similar time-scale. Its second brightening episode which triggered the \textit{Gaia} alerts system had an amplitude of $\sim$2.5 mag.  

\item[-] We have taken optical and NIR spectra of the sources using the TNG during their bright state, and NIR spectra for Gaia20fgx using the GTC during its fading. The hydrogen Balmer lines from H$\alpha$ to H$\delta$, Pa$\beta$, Br$\gamma$, and lines of \ion{Ca}{ii}, \ion{O}{i}, and \ion{Na}{i} were detected in emission in the spectra of Gaia20bwa. The H$\alpha$ and Br$\gamma$ lines were detected toward Gaia20fgx in emission in its bright state, with additional 2-0 and 3-1 bandhead features of CO in absorption and the Pa$\beta$ line showing an inverse P Cygni profile during its fading.

\item[-] Based on the Br$\gamma$ lines the accretion rate was found to be $(2.4-3.1)\times10^{-8}$ $M_\odot$ yr$^{-1}$ for Gaia20bwa and $(4.5-6.6)\times10^{-8}$ $M_\odot$ yr$^{-1}$ for Gaia20fgx during their bright state. The accretion rate of Gaia20fgx dropped by about a factor of 10, to $(6.01 - 7.0) \times 10^{-9}$ $M_\odot$ year$^{-1}$, on a time-scale of half a year.

\item[-] The accretion luminosities and rates measured for both sources are closer to those found for CTTS for similar stellar luminosities and masses than to those measured for young eruptive stars. However, the amplitude and time-scale of these brightening events place these two stars to a region of the parameter space, which is rarely populated by accreting young stellar objects. This suggests a new class of young stellar objects, which produce outbursts on a time-scale similar to young eruptive stars, but with smaller amplitudes, possibly representing an intermediate case between variable CTTS and young eruptive stars. 
    
\end{itemize}

\section*{Acknowledgements}

We thank the referee for comments, which helped to improve our paper.
This project has received funding from the European Research Council (ERC) under the European Union's Horizon 2020 research and innovation programme under grant agreement No 716155 (SACCRED).
We acknowledge support from the ESA PRODEX contract nr. 4000132054. \\
GM acknowledges funding from the European Union’s Horizon 2020 research and innovation programme under grant agreement No 101004141. \\
PZ and LW acknowledges the support from  European Commission's H2020 OPTICON grant No. 730890, OPTICON RadioNet Pilot grant No. 101004719 as well as Polish NCN grant Daina No. 2017/27/L/ST9/03221. \\ 
AP acknowledges support from Grant K-138962 of the National Research, Development and Innovation Office (NKFIH, Hungary). \\
LK acknowledges the financial support of the Hungarian National Research, Development and Innovation Office grant NKFIH PD-134784 and K-131508. LK is a Bolyai J\'anos Research Fellow. \\
Based on observations made with the Italian Telescopio Nazionale Galileo (TNG) operated by the Fundación Galileo Galilei (FGG) of the Istituto Nazionale di Astrofisica (INAF) at the Observatorio del Roque de los Muchachos (La Palma, Canary Islands, Spain). \\
SA, TG and BN received financial support from the project PRIN-INAF 2019 "Spectroscopically Tracing the Disk Dispersal Evolution"  and by  the  project PRIN-INAF-MAIN-STREAM 2017  “Protoplanetary    disks    seen through  the  eyes  of  new-generation  instruments”. \\
We acknowledge Ennio Poretti and Avet Harutyunyan for their help as TNG experts. \\
The Liverpool Telescope is operated on the island of La Palma by Liverpool John Moores University in the Spanish Observatorio del Roque de los Muchachos of the Instituto de Astrofisica de Canarias with financial support from the UK Science and Technology Facilities Council. 
\\
Based on observations made with the Gran Telescopio Canarias (GTC), installed in the Spanish Observatorio del Roque de los Muchachos of the Instituto de Astrofisica de Canarias, in the island of La Palma. \\
This work is (partly) based on data obtained with the instrument EMIR, built by a Consortium led by the Instituto de Astrofisica de Canarias. EMIR was funded by GRANTECAN and the National Plan of Astronomy and Astrophysics of the Spanish Government.

\section*{Data availability}

The data underlying this article will be shared on reasonable request to the corresponding author.

\bibliographystyle{mnras}
\bibliography{gaiasources}

%%%%%%%%%%%%%%%%%%%%%%%%%%%%%%%%%%%%%%%%%%%%%%%%%%

%%%%%%%%%%%%%%%%% APPENDICES %%%%%%%%%%%%%%%%%%%%%

\appendix

\section{Photometry}
\label{sec:appendix}

\begin{table*}
\centering
\scriptsize
\caption{Optical and near-infrared photometry of Gaia20bwa and Gaia20fgx. Uncertainties are typically 0.1 mag in the $B$ filter, 0.05 mag in the $gVri$ filters and 0.01 mag in the $JHK_s$ filters.}\label{tab:phot}
\begin{tabular}{cccccccccccc}
\hline \hline
Date       & JD$\,{-}\,$2\,450\,000 & $B$ & $g$ & $V$ & $r$ & $i$ & $J$ & $H$ & $K_s$ & Telescope \\
\hline
\multicolumn{11}{c}{Gaia20bwa}\\
\hline
2021-01-03 & 9218.36 & \dots    & 17.607     & \dots       & 16.701     & 15.507     & \dots      & \dots      & \dots      & Mt.~Suhora \\
2021-01-22 & 9237.35 & \dots    & 17.595     & \dots       & 16.646     & 15.451     & \dots      & \dots      & \dots      & Mt.~Suhora \\
2021-02-11 & 9257.41 & \dots    & \dots      & \dots       & \dots      & \dots      & 12.930     & 12.164     & 11.741     & TNG \\
2021-02-12 & 9258.31 & 18.147   & \dots      & 17.315      & 16.712     & 15.580     & \dots      & \dots      & \dots      & RC80 \\
2021-02-13 & 9259.29 & 18.004   & \dots      & 17.073      & 16.596     & 15.456     & \dots      & \dots      & \dots      & RC80 \\
2021-03-20 & 9294.30 & 17.462   & \dots      & 16.865      & 16.458     & 15.454     & \dots      & \dots      & \dots      & RC80 \\
2021-09-02 & 9459.58 & \dots    & \dots      & \dots       & \dots      & 15.869     & \dots      & \dots      & \dots      & RC80 \\
2021-09-07 & 9464.60 & 19.291   & \dots      & 18.218      & 17.496     & 15.805     & \dots      & \dots      & \dots      & RC80 \\
2021-09-08 & 9465.58 & \dots    & \dots      & 17.974      & 17.414     & 15.694     & \dots      & \dots      & \dots      & RC80 \\
2021-09-09 & 9466.57 & \dots    & \dots      & 18.267      & 17.521     & 15.823     & \dots      & \dots      & \dots      & RC80 \\
2021-09-10 & 9467.58 & 19.249   & \dots      & 18.069      & 17.376     & 15.741     & \dots      & \dots      & \dots      & RC80 \\
2021-09-16 & 9473.63 & \dots    & \dots      & 18.117      & 17.824     & 15.914     & \dots      & \dots      & \dots      & RC80 \\
2021-09-26 & 9483.53 & 19.068   & \dots      & 18.052      & 17.406     & 15.774     & \dots      & \dots      & \dots      & RC80 \\
2021-09-27 & 9484.57 & \dots    & \dots      & 18.120      & 17.718     & 15.872     & \dots      & \dots      & \dots      & RC80 \\
2021-10-04 & 9491.53 & \dots    & \dots      & 18.257      & 17.540     & 15.861     & \dots      & \dots      & \dots      & RC80 \\
2021-10-15 & 9502.63 & 19.298   & \dots      & 18.199      & 17.471     & 15.713     & \dots      & \dots      & \dots      & RC80 \\
2021-10-20 & 9507.59 & \dots    & \dots      & 18.296      & 17.564     & 15.815     & \dots      & \dots      & \dots      & RC80 \\
2021-10-24 & 9512.49 & \dots    & \dots      & 18.191      & 17.403     & 15.749     & \dots      & \dots      & \dots      & RC80 \\
2021-10-30 & 9517.60 & 19.484   & \dots      & 18.295      & 17.495     & 15.811     & \dots      & \dots      & \dots      & RC80 \\
2021-11-12 & 9530.59 & 19.498   & \dots      & 18.182      & 17.504     & 15.806     & \dots      & \dots      & \dots      & RC80 \\
2021-12-14 & 9562.56 & 19.371   & \dots      & 18.219      & 17.499     & 15.821     & \dots      & \dots      & \dots      & RC80 \\
2022-01-12 & 9592.39 & \dots  & \dots & 18.232 & 17.557 & 15.843 & \dots & \dots & \dots & RC80 \\
2022-01-15 & 9595.33 & \dots  & \dots & \dots  & \dots  & 15.809 & \dots & \dots & \dots & RC80 \\
2022-02-23 & 9634.26 & 19.397 & \dots & 18.263 & 17.517 & 15.834 & \dots & \dots & \dots & RC80 \\
\hline
\multicolumn{11}{c}{Gaia20fgx}\\
\hline
2021-01-22 & 9237.28 & \dots  & 19.629      & \dots       & 17.523     & 16.306     & \dots       & \dots      & \dots      & Mt.~Suhora \\
2021-01-27 & 9242.34 & \dots  & \dots       & \dots       & \dots      & \dots      & 13.777      & 12.661     & 12.039     & TNG \\
2021-02-13 & 9259.24 & \dots  & \dots       & 18.827      & 17.905     & 16.877     & \dots       & \dots      & \dots      & RC80 \\
2021-02-14 & 9260.30 & \dots  & \dots       & 18.694      & 17.871     & 16.748     & \dots       & \dots      & \dots      & RC80 \\
2021-05-05 & 9339.51 & \dots  & \dots       & 19.152      & 18.418     & 17.307     & \dots       & \dots      & \dots      & RC80 \\
2021-05-09 & 9343.56 & \dots  & \dots       & 19.121      & 18.245     & 17.144     & \dots       & \dots      & \dots      & RC80 \\
2021-06-19 & 9384.52 & \dots  & \dots       & 19.679      & 18.752     & 17.553     & \dots       & \dots      & \dots      & RC80 \\
2021-07-06 & 9401.54 & \dots  & \dots       & 19.649      & 18.536     & 17.370     & \dots       & \dots      & \dots      & RC80 \\
2021-07-09 & 9404.55 & \dots  & \dots       & 19.432      & 18.496     & 17.538     & \dots       & \dots      & \dots      & RC80 \\
2021-07-12 & 9408.50 & \dots  & \dots       & 19.646      & 18.697     & 17.588     & \dots       & \dots      & \dots      & RC80 \\
2021-07-24 & 9420.50 & \dots  & 20.133      & \dots       & 18.536     & 17.211     & \dots       & \dots      & \dots      & Mt.~Suhora \\
2021-07-30 & 9425.51 & \dots  & \dots       & 19.204      & 18.585     & 17.688     & \dots       & \dots      & \dots      & RC80 \\
2021-07-30 & 9426.45 & \dots  & \dots       & 19.750      & 18.736     & 17.676     & \dots       & \dots      & \dots      & RC80 \\
2021-07-31 & 9427.45 & \dots  & \dots       & 19.641      & 18.671     & 17.634     & \dots       & \dots      & \dots      & RC80 \\
2021-08-10 & 9437.45 & \dots  & \dots       & 19.089      & 18.119     & 17.049     & \dots       & \dots      & \dots      & RC80 \\
2021-08-11 & 9438.49 & \dots  & \dots       & 19.212      & 18.169     & 16.986     & \dots       & \dots      & \dots      & RC80 \\
2021-08-13 & 9439.50 & \dots  & \dots       & 19.257      & 18.272     & 17.062     & \dots       & \dots      & \dots      & RC80 \\
2021-08-14 & 9440.54 & \dots  & \dots       & 19.462      & 18.523     & 17.329     & \dots       & \dots      & \dots      & RC80 \\
2021-08-14 & 9441.45 & \dots  & \dots       & 19.128      & 18.226     & 17.197     & \dots       & \dots      & \dots      & RC80 \\
2021-08-15 & 9441.52 & \dots  & 20.421      & \dots       & 18.526     & 17.248     & \dots       & \dots      & \dots      & Mt.~Suhora \\
2021-08-18 & 9445.48 & \dots  & 20.163      & \dots       & 18.706     & 17.468     & \dots       & \dots      & \dots      & Mt.~Suhora \\
2021-08-21 & 9447.53 & \dots  & \dots       & 19.535      & 18.444     & 17.238     & \dots       & \dots      & \dots      & RC80 \\
2021-08-28 & 9454.60 & \dots  & \dots       & 19.457      & 18.538     & 17.329     & \dots       & \dots      & \dots      & RC80 \\
2021-08-31 & 9457.61 & \dots  & \dots       & 20.143      & 18.682     & 17.382     & \dots       & \dots      & \dots      & RC80 \\
2021-09-05 & 9463.42 & \dots  & \dots       & \dots       & 18.098     & 17.164     & \dots       & \dots      & \dots      & RC80 \\
2021-09-06 & 9464.41 & \dots  & \dots       & 19.142      & 18.227     & 17.083     & \dots       & \dots      & \dots      & RC80 \\
2021-09-07 & 9465.37 & \dots  & \dots       & 19.039      & 18.152     & 16.981     & \dots       & \dots      & \dots      & RC80 \\
2021-09-09 & 9466.51 & \dots  & \dots       & 18.876      & 18.058     & 16.975     & \dots       & \dots      & \dots      & RC80 \\
2021-09-09 & 9467.47 & \dots  & \dots       & 19.595      & 18.555     & 17.318     & \dots       & \dots      & \dots      & RC80 \\
2021-09-10 & 9468.38 & \dots  & \dots       & 19.055      & 18.279     & 17.228     & \dots       & \dots      & \dots      & RC80 \\
2021-09-11 & 9469.36 & \dots  & \dots       & 19.664      & 18.613     & 17.505     & \dots       & \dots      & \dots      & RC80 \\
2021-09-12 & 9470.44 & \dots  & \dots       & 19.463      & 18.579     & 17.403     & \dots       & \dots      & \dots      & RC80 \\
2021-09-13 & 9471.34 & \dots  & \dots       & 19.514      & 18.555     & 17.354     & \dots       & \dots      & \dots      & RC80 \\
2021-09-18 & 9476.43 & \dots  & \dots       & \dots       & \dots      & \dots      & 14.060      & 12.856     & 12.229     & GTC  \\
2021-09-26 & 9483.50 & \dots  & \dots       & 19.326      & 18.280     & 17.012     & \dots       & \dots      & \dots      & RC80 \\
2021-09-27 & 9484.59 & \dots  & \dots       & 19.252      & 18.151     & 16.867     & \dots       & \dots      & \dots      & RC80 \\
2021-10-24 & 9511.58 & \dots  & \dots       & 19.560      & 18.514     & 17.481     & \dots       & \dots      & \dots      & RC80 \\
2021-10-26 & 9513.58 & \dots  & \dots       & \dots       & 18.303     & 17.643     & \dots       & \dots      & \dots      & RC80 \\
2021-10-28 & 9515.57 & \dots  & \dots       & 19.926      & 18.914     & 17.812     & \dots       & \dots      & \dots      & RC80 \\
2021-10-30 & 9517.56 & \dots  & \dots       & \dots       & 18.893     & 17.920     & \dots       & \dots      & \dots      & RC80 \\
2021-11-01 & 9519.58 & \dots  & \dots       & \dots       & 18.376     & 17.564     & \dots       & \dots      & \dots      & RC80 \\
2021-11-09 & 9527.56 & \dots  & \dots       & 19.579      & 18.900     & 17.752     & \dots       & \dots      & \dots      & RC80 \\
2021-11-12 & 9530.52 & \dots  & \dots       & 19.744      & 19.177     & 18.137     & \dots       & \dots      & \dots      & RC80 \\
2021-11-15 & 9533.55 & \dots  & \dots       & \dots       & 19.092     & 18.390     & \dots       & \dots      & \dots      & RC80 \\
2021-11-17 & 9535.52 & \dots  & \dots       & 20.236      & 19.281     & 18.249     & \dots       & \dots      & \dots      & RC80 \\
2021-11-24 & 9542.54 & \dots  & \dots       & \dots       & 19.031     & 18.034     & \dots       & \dots      & \dots      & RC80 \\
2021-11-30 & 9548.52 & \dots  & \dots       & \dots       & 18.954     & 18.139     & \dots       & \dots      & \dots      & RC80 \\
2021-12-04 & 9552.51 & \dots  & \dots       & \dots       & 19.016     & 18.424     & \dots       & \dots      & \dots      & RC80 \\
2021-12-08 & 9556.51 & \dots  & \dots       & \dots       & 18.972     & 18.126     & \dots       & \dots      & \dots      & RC80 \\
2021-12-13 & 9562.40 & \dots  & \dots       & 19.748      & 18.940     & 17.930     & \dots       & \dots      & \dots      & RC80 \\
2021-12-17 & 9566.45 & \dots  & \dots       & \dots       & 18.882     & 18.055     & \dots       & \dots      & \dots      & RC80 \\
2022-01-01 & 9581.44 & \dots  & \dots       & 19.262      & 18.355     & 17.175     & \dots       & \dots      & \dots      & RC80 \\
2022-02-02 & 9613.69 & \dots  & \dots & 19.019 & 18.382 & 17.433 & \dots & \dots & \dots & RC80 \\
\hline
\end{tabular}
\end{table*}

\end{document}